\documentclass[aps,prd,groupedaddress,showpacs,showkeys]{revtex4}
\usepackage{amssymb}
\usepackage{amsmath}
\usepackage{amsfonts}
\usepackage{epsfig}
\usepackage{amssymb}
\usepackage{amsmath}
\usepackage{amsfonts}
\usepackage{bm}
\usepackage{epsfig}

\def\shiftdown#1{#1\llap{\lower.04ex\hbox{#1}}}

\newcommand{\seq}{\begin{subequations}}
\newcommand{\sen}{\end{subequations}}
\newcommand{\eq}{\begin{eqnarray}}
\newcommand{\en}{\end{eqnarray}}

\newcommand{\la}{\langle} 
\newcommand{\ra}{\rangle} 
\newcommand{\beqn}{\begin{eqnarray}}
\newcommand{\eeqn}{\end{eqnarray}}
\newcommand{\beq}{\begin{equation}}
\newcommand{\eeq}{\end{equation}}
\newcommand{\barr}{\begin{array}}
\newcommand{\earr}{\end{array}}

\begin{document}

\title{Radiative decays of double heavy baryons in a relativistic \\ 
       constituent three--quark model including hyperfine mixing
effects} 
\noindent
\author{Tanja Branz$^{1}$, 
        Amand Faessler$^{1}$,
        Thomas Gutsche$^{1}$,
        Mikhail~A.~Ivanov$^{2}$,  
        J\"urgen G. K\"{o}rner$^{3}$, 
        Valery~E.~Lyubovitskij$^1$\footnote{On leave of absence
          from Department of Physics, Tomsk State University,
          634050 Tomsk, Russia}, 
        Bettina Oexl$^{1}$  
\vspace*{1.2\baselineskip}}

\affiliation{$^1$ Institut f\"ur Theoretische Physik,
Universit\"at T\"ubingen,\\ 
Kepler Center for Astro and Particle Physics, \\
Auf der Morgenstelle 14, D--72076 T\"ubingen, Germany
\vspace*{1.2\baselineskip} \\
$^2$ Bogoliubov Laboratory of Theoretical Physics,
Joint Institute for Nuclear Research,~141980~Dubna,~Russia
\vspace*{1.2\baselineskip} \\
$^3$ Institut f\"{u}r Physik, Johannes Gutenberg-Universit\"{a}t,
D--55099 Mainz, Germany\\}

\date{\today}
\begin{abstract}

We study flavor--conserving radiative decays of double heavy baryons using 
a manifestly Lorentz covariant constituent three--quark model. 
Decay rates are calculated and compared to each other in the full theory, 
keeping masses finite, and also in the heavy quark limit. We discuss in some
detail hyperfine mixing effects. 

\end{abstract}



\maketitle

\newpage

\section{Introduction}

A first observation of the double charmed baryon $\Xi_{cc}^+(3519)$ 
by the SELEX Collaboration at Fermilab~\cite{Mattson:2002vu} 
stimulated theoretical studies of double heavy baryons (DHBs). 
Up to now the study of DHBs has mainly focussed on 
their mass spectra and their semileptonic decays (for an overview see e.g. 
Refs.~\cite{Kiselev:2001fw,Faessler:2009xn}). 
In particular, significant progress has been achieved in the analysis 
of the DHB semileptonic weak decays. The current--induced flavor-changing 
double--heavy baryon  transitions
have been analyzed in a number of model approaches. These include  
effective field theories based on heavy quark spin 
symmetry~\cite{White:1991hz,SanchisLozano:1994vh,%
Flynn:2007qt,Hernandez:2007qv}, 
three--quark models~\cite{Faessler:2001mr,Albertus:2006ya,%
Roberts:2008wq,Albertus:2009ww}, 
quark--diquark models~\cite{Guo:1998yj,Ebert:2004ck}, and 
nonrelativistic QCD sum rules~\cite{Onishchenko:2000wf,Kiselev:2001fw}. 
Recently~\cite{Faessler:2009xn} we have presented a comprehensive analysis 
of the semileptonic decays of DHBs using a manifestly Lorentz covariant 
field theory approach termed the relativistic constituent three--quark model 
(RTQM)~\cite{Faessler:2009xn,Faessler:2001mr,Ivanov:1996pz}. 
We considered all possible current--induced spin transitions  
between double--heavy baryons containing both types of light quarks -- 
nonstrange $q=u,d$ and strange $s$. These involved the flavor-changing 
transitions
$bc \to cc$ and $bb \to bc$. Form factors and decay rates have been 
calculated and have been compared to each other in the full theory with 
all masses finite, and also in the heavy quark limit (HQL). Such an analysis 
is important because the semileptonic decays of DHBs provide yet another 
opportunity to measure the Cabibbo--Kobayashi--Maskawa (CKM) matrix element 
$V_{cb}$. This is particularly true since the transition matrix elements 
between
double--heavy baryons obey spin symmetry relations in the heavy quark limit 
in addition to a model independent zero recoil normalization of the relevant 
transition matrix elements. 

In this paper we continue the study of DHB properties in the 
RTQM~\cite{Faessler:2009xn,Ivanov:1996pz}. In particular, 
we analyze flavor--conserving radiative transitions between ground state DHBs: 
$1/2^+ \to 1/2^+$ and $3/2^+ \to 1/2^+$. The first estimate 
of the radiative decay widths of the DHBs 
in the heavy quark limit including hyperfine mixing effects
has been done in Ref.~\cite{Albertus:2009ww} 
(for radiative transitions between DHBs see also the early paper 
\cite{Dai:2000hza}). A detailed analysis of DHB decays containing only the 
$(bc)$ heavy quark configuration has been considered 
recently in Ref.~\cite{Albertus:2010hi}. As in our recent 
paper~\cite{Faessler:2009xn}, we take the DHBs to be bound 
states of a light quark and a double--heavy ($Q_1Q_2$) diquark. 

The origin of the hyperfine mixing for double heavy baryons is the one--gluon
exchange interaction between the light and heavy quarks in the DHB states
containing two different heavy quarks --- $b$ and $c$. It leads to
mixing of the states containing spin--0 and spin--1 heavy
quark configurations.
As shown in Refs.~\cite{Roberts:2008wq,Albertus:2009ww,Albertus:2010hi}
hyperfine mixing has a big impact on the decay properties of double heavy
baryons. Both the weak semileptonic and the electromagnetic decay rates
involving mixed DHB states are significantly enhanced or reduced relative to 
the rates involving unmixed states. 

The RTQM can be viewed as an effective 
quantum field theory approach based on an interaction Lagrangian of hadrons  
interacting with their constituent quarks. From such an approach one can
derive universal and reliable predictions 
for exclusive processes involving both mesons composed of a quark and 
antiquark and baryons composed of three quarks. 
The coupling strength of a hadron $H$  to its constituent quarks
is determined by the compositeness condition 
$Z_H=0$~\cite{Weinberg:1962hj,Efimov:1993ei} where 
$Z_H$ is the wave function renormalization constant of the hadron $H$. The 
quantity $Z_H^{1/2}$ is the matrix element between the physical particle state 
and the corresponding bare state. The compositeness condition $Z_H=0$ enables 
one to represent a bound state by introducing a hadronic field interacting 
with its constituents so that the renormalization factor is equal to zero. 
This does not mean that we can solve the QCD bound state equations but we are 
able to show that the condition $Z_H=0$ provides an effective and 
self--consistent way to describe the coupling of a hadron to its 
constituents. One  starts with an effective interaction Lagrangian written 
down in terms of quark and hadron variables. Then, by using Feynman rules, 
the $S$--matrix elements describing hadron-hadron interactions are given in 
terms of a set of quark level Feynman diagrams. In particular, the 
compositeness condition enables one to avoid the problem of double counting of
quark and hadronic degrees of 
freedom. The approach is self--consistent and all calculations of physical 
observables are straight--forward. There is  a small set of  model parameters: 
the values of the constituent quark masses and the scale parameters that 
define the size of the distribution of the constituent quarks inside a given 
hadron. 

The main objective of the present paper is to present 
an analysis of all possible electromagnetic transitions 
between ground state DHBs containing both types of light quarks -- 
nonstrange $q=u,d$ and strange $s$. 
The paper is structured as follows. 
First, in Sec.II we review our relativistic constituent three--quark model
approach (for more details see e.g.~\cite{Faessler:2009xn}) including a 
discussion on how to obtain
a gauge invariant coupling of the photon in our model. In Sec.III we discuss 
in more detail various aspects of the radiative decays of DHBs. We
discuss the calculation of the relevant radiative transition matrix elements
and analyze the consequences of taking the heavy quark limit for the
radiative transitions. In Sec.IV we discuss in some detail hyperfine mixing 
effects in the radiative decays of DHBs. Sec.V contains our numerical results 
which are compared to the predictions of a naive nonrelativistic quark model 
that has the same spin--flavor symmetry group as our DHB currents in the 
nonrelativistic limit. We also compare the results of the full finite 
mass calculation with results derived in the HQL. In addition we 
compare our results for $bc \to bc$ radiative transitions 
with recent quark model results~\cite{Albertus:2010hi}.  
Finally, in Sec.~VI we present a brief summary of our results.

\section{Framework}  

\subsection{Lagrangian}

For the evaluation of the radiative decays of DHBs 
we will consistently employ the RTQM~\cite{Faessler:2009xn,Ivanov:1996pz}.  
The model is based on an interaction Lagrangian describing 
the coupling between a baryon $B(q_1q_2q_3)$ and its constituent 
quarks $q_1$, $q_2 $ and $q_3$. For $J^P= \frac{1}{2}^+$ and 
$\frac{3}{2}^+$ baryons the Lagrangians read: 
\eq\label{Lagr_str}
{\cal L}_{\rm int}^{\rm str}(x) &=& g_B \bar B(x) \, 
\int\!\! dx_1 \!\! \int\!\! dx_2 \!\! \int\!\! dx_3 \, 
F(x,x_1,x_2,x_3) \, J_B(x_1,x_2,x_3) \nonumber\\ 
&+& g_{B^\ast} \bar B^{\ast}_\mu(x) \, 
\int\!\! dx_1 \!\! \int\!\! dx_2 \!\! \int\!\! dx_3 \, 
F(x,x_1,x_2,x_3) \, J_{B^\ast}^\mu(x_1,x_2,x_3) \, + \, {\rm h.c.}  
\en 
where $J_{B}$ and $J_{B^\ast}$ are interpolating three--quark currents 
with the quantum numbers of the relevant baryon $B(\frac{1}{2}^+)$ and 
$B^\ast(\frac{3}{2}^+)$. Note that the spin $3/2$ spinor corresponding to 
the $B^\ast(\frac{3}{2}^+)$ field satisfies subsidiary
Rarita--Schwinger conditions (see further details in Appendix A).  

One has
\eq 
J_{B}(x_1,x_2,x_3) \, = \, \varepsilon^{a_1a_2a_3} \, 
\Gamma_1 \, q^{a_3}(x_3) \, Q^{a_1}_1(x_1) 
C \, \Gamma_2 \, Q^{a_2}_2(x_2) \, , \\
J_{B^\ast}^\mu(x_1,x_2,x_3) \, = \, \varepsilon^{a_1a_2a_3} \, 
\Gamma_1 \, q^{a_3}(x_3) \, Q^{a_1}_1(x_1) 
C \, \Gamma_2^\mu \, Q^{a_2}_2(x_2) \, ,
\label{lag}
\en 
where the $\Gamma_{1,2}$ are strings of Dirac matrices, $C$ is the charge 
conjugation 
matrix $C=\gamma^{0}\gamma^{2}$ and the $a_i$ ($i$=1,2,3) are color indices.
$F(x,x_1,x_2,x_3)$ is a nonlocal scalar vertex function which characterizes
the finite size of the baryons. 

The full Lagrangian 
\eq\label{Lagr_full}
{\cal L}_{\rm full}(x) \, = \, {\cal L}_{\rm free}(x) \, + \, 
{\cal L}_{\rm int}^{\rm em(1)}(x)  \, + \, 
{\cal L}_{\rm int}^{\rm str + em(2)}(x)  
\en 
needed for the calculation of the 
radiative decays of DHBs includes the free parts of the baryons and the 
constituent quarks 
\eq 
{\cal L}_{\rm free}(x) \, = \, \bar B(x) \, {\cal D}_B \, B(x) 
\, - \, \bar B^{\ast}_\mu(x) \, {\cal D}_{B^\ast}^{\mu\nu} \, B^\ast_{\nu} 
\, + \, \sum\limits_{\psi=q,Q} \, \bar \psi(x) \, 
{\cal D}_{\psi} \, \psi(x) \, ,
\en 
where 
\eq 
{\cal D}_{\psi} &=& i\not\! \partial - m_{\psi} \,, \nonumber\\
{\cal D}_{B} &=& i\not\! \partial - m_{B} \,, \nonumber\\
{\cal D}_{B^\ast}^{\mu\nu} &=& g^{\mu\nu} ( i \not\! \partial - m_{B^\ast}) 
- i ( \gamma^\mu \partial^\nu + \gamma^\nu \partial^\mu ) 
+ \gamma^\mu i \not\! \partial \gamma^\nu - m_{B^\ast} 
\gamma^\mu \gamma^\nu \,. 
\en
The baryon and constituent quark masses are denoted by 
$m_{B(B^\ast)}$ and $m_{\psi}$, 
respectively.

The electromagnetic interaction Lagrangian contains two pieces given by 
\eq 
{\cal L}_{\rm int}^{\rm em} = 
{\cal L}^{\rm em (1)}_{\rm int} + {\cal L}^{\rm em (2)}_{\rm int}
\en 
which are generated after the inclusion of photons. 
The first term ${\cal L}^{\rm em (1)}_{\rm int}$ 
is generated via minimal substitution in the free Lagrangian 
${\cal L}_{\rm free}$:
\eq
\label{photon}
\partial^\mu \Psi \to (\partial^\mu - ie_\Psi A^\mu) \Psi\,,
\hspace*{.5cm}
\partial^\mu \overline\Psi \to (\partial^\mu + ie_\Psi A^\mu)\overline\Psi\,,
\en 
where $\Psi$ stands for $ B, B^\ast, q, Q$, and 
$e_\Psi$ is the electric charge of the field $\Psi$.  
The interaction Lagrangian ${\cal L}^{\rm em (1)}_{\rm int}$ reads 
\eq
\label{L_em_1}
{\cal L}^{\rm em (1)}_{\rm int}(x) =
  e_B \bar B(x) \!\not\!\! A \, B(x)  
- e_{B^\ast} \bar B^\ast_\mu(x) 
\biggl( \, \!\not\!\! A \ g^{\mu\nu} 
+ \gamma^\mu \!\not\!\! A \ \gamma^\nu 
- \gamma^\mu A^\nu - \gamma^\nu A^\mu \biggr) B^\ast_\nu(x) 
+ \sum\limits_{\psi=q,Q} \, e_\psi \, \bar\psi(x) \!\not\!\! A  \, \psi(x) \,.
\en
The second electromagnetic interaction term ${\cal L}^{\rm em (2)}_{\rm int}$
is generated when one gauges the nonlocal 
Lagrangian Eq.~(\ref{Lagr_str}). 
The gauging proceeds in a way suggested and extensively used
in Refs.~\cite{Ivanov:1996pz,Mandelstam:1962mi,Terning:1991yt}.
In order to guarantee local electromagnetic gauge invariance of the strong 
interaction
Lagrangian one multiplies
each quark field in ${\cal L}_{\rm int}^{\rm str}$ with a
gauge field exponential. 
One then has 
\eq\label{gauging}
{\cal L}_{\rm int}^{\rm str + em(2)}(x) &=& g_B \bar B(x) \,
\int\!\! dx_1 \!\! \int\!\! dx_2 \!\! \int\!\! dx_3 \,
F(x,x_1,x_2,x_3) \, \epsilon^{a_1a_2a_3} \, \Gamma_1
e^{-ie_{q} I(x_3,x,P)} \, q^{a_3}(x_3) \nonumber\\
&\times& \, e^{-ie_{Q_1} I(x_1,x,P)} \, Q^{a_1}_1(x_1)
C \, \Gamma_2 \,
\, e^{-ie_{Q_2} I(x_2,x,P)} \, Q^{a_2}_2(x_2) \,,  \nonumber\\ 
&+&g_{B^\ast} \bar B^\ast_\mu(x) \,
\int\!\! dx_1 \!\! \int\!\! dx_2 \!\! \int\!\! dx_3 \,
F(x,x_1,x_2,x_3) \, \epsilon^{a_1a_2a_3} \, \Gamma_1
e^{-ie_{q} I(x_3,x,P)} \, q^{a_3}(x_3) \nonumber\\
&\times& \, e^{-ie_{Q_1} I(x_1,x,P)} \, Q^{a_1}_1(x_1)
C \, \Gamma_2^\mu \,
\, e^{-ie_{Q_2} I(x_2,x,P)} \, Q^{a_2}_2(x_2)
\, + \, {\rm H.c.}  
\en
where
\eq\label{path}
I(x_i,x,P) = \int\limits_x^{x_i} dz_\mu A^\mu(z).
\en
An expansion of the gauge exponential up to a certain power of $A^\mu$
leads to the terms contained in ${\cal L}^{\rm em (2)}_{\rm int}$.

The full Lagrangian consistently generates all the required matrix elements 
of the radiative decays of the DHBs. The relevant transitions can be 
represented by a set of 
quark loop 
diagrams. In the evaluation of the quark loops we use the free fermion 
propagator for the constituent quarks as dictated by the free quark Lagrangian 
discussed above. One has 
\eq\label{quark_propagator}
i \, S_{\psi}(x-y) = \langle 0 | T \, \psi(x) \, \bar \psi(y)  | 0 \rangle
= \int\frac{d^4k}{(2\pi)^4i} \, e^{-ik(x-y)} \ \tilde S_{\psi}(k)
\en
where
\eq
\tilde S_{\psi}(k) = \frac{1}{m_{\psi}-\not\! k -i\epsilon}
\en
is the usual free fermion propagator in momentum space.
We avoid the appearance of unphysical imaginary parts
in Feynman diagrams by postulating that
the baryon mass is less than the sum of the constituent
quark masses $m_{B(B^\ast)} < m_{q}+m_{Q_1}+m_{Q_2}$ which is satisfied
in our calculation.
We mention that we have recently introduced a further 
refinement of our model in that we can now include
quark confinement effects ~\cite{Branz:2009cd}.

The free propagators of the baryon fields in momentum space are given by 
\eq 
\tilde S_B(k) &=& \frac{1}{m_B-\not\! k -i\epsilon} \,, \\
\tilde S_{B^\ast}^{\mu\nu}(k) &=& 
\frac{1}{m_{B^\ast}-\not\! k -i\epsilon} \, 
\biggl( - g^{\mu\nu} + \frac{1}{3} \gamma^\mu \gamma^\nu 
+ \frac{2 \, k^\mu k^\nu}{3 \, m_{B^\ast}^2} 
+ \frac{k^\nu \gamma^\mu - k^\mu \gamma^\nu}{3 \, m_{B^\ast}} 
\biggr) \,. 
\en
Next we consider in detail the required building blocks of the strong 
interaction Lagrangian $ {\cal L}_{\rm int}^{\rm str}$ --- 
the vertex function $F$, the interpolating three--quark currents 
$J_{B}$ and $J_{B^\ast}^\mu$,  
and the baryon-quark coupling constants $g_B$ and $g_{B^\ast}$. 

\subsection{Vertex function}

The vertex function $F_B$ is related to the scalar part of the 
Bethe--Salpeter amplitude and characterizes the finite size 
of the baryon. In our approach we use a specific form for 
the vertex function given by  
\eq\label{vertex}
F(x,x_1,x_2,x_3) \, = \, N \, 
\delta^{(4)}(x - \sum\limits_{i=1}^3 w_i x_i) \;  
\Phi\biggl(\sum_{i<j}( x_i - x_j )^2 \biggr) \,, 
\en 
which is Poincar\'e--invariant.  
$\Phi$ is a nonlocal correlation function involving the three 
constituent quarks with masses $m_1$, $m_2$, $m_3$; $N = 9$ is a 
normalization factor. The variable $w_i$ is defined by 
$w_i=m_i/(m_1+m_2+m_3)$. 

The Fourier transform of the correlation function
$\Phi\biggl(\sum\limits_{i<j} ( x_i - x_j )^2 \biggr)$ can be calculated
by using Jacobi coordinates. One has
\begin{eqnarray}
\label{fourier}
\Phi(p_1,p_2,p_3) &=& N \int dx e^{-ipx} 
\prod\limits_{i=1}^3\int\! dx_i e^{ip_ix_i}\, 
\delta^{(4)}(x - \sum\limits_{i=1}^3 w_i x_i) \, 
\Phi\biggl( \sum_{i<j}( x_i - x_j )^2 \biggr)  
\nonumber\\
&=& (2\pi)^4\,\delta^{(4)}\Big(p  - \sum\limits_{i=1}^3 p_i\Big)\,
\Phi(-l_1^2-l_2^2)\,,
\end{eqnarray}
where the  Jacobi coordinates are defined by  
\begin{eqnarray}
x_1&=&x \, + \, \tfrac{1}{\sqrt{2}}\, \xi_1 w_3 
        \, - \, \tfrac{1}{\sqrt{6}}\, \xi_2 (2 w_2 + w_3)\,,
\nonumber\\
x_2&=&x \, + \, \tfrac{1}{\sqrt{2}}\, \xi_1 w_3      
        \, + \, \tfrac{1}{\sqrt{6}}\, \xi_2 (2 w_1 + w_3)\,,
\nonumber\\
x_3&=&x \, - \, \tfrac{1}{\sqrt{2}}\, \xi_1  (w_1  + w_2)     
        \, + \, \tfrac{1}{\sqrt{6}}\,  \xi_2 (w_1  - w_2)\,. 
\end{eqnarray}
The corresponding Jacobi momenta read
\begin{eqnarray}
p &=& p_1+p_2+p_3\,,
\nonumber\\
l_1 &=& \tfrac{1}{\sqrt{2}}\,w_3 (p_1 + p_2) 
      - \tfrac{1}{\sqrt{2}}\,(w_1+w_2) p_3\,, 
\nonumber\\
l_2 &=&  - \tfrac{1}{\sqrt{6}}\,(2 w_2 + w_3) p_1
         + \tfrac{1}{\sqrt{6}}\,(2 w_1 + w_3) p_2
         + \tfrac{1}{\sqrt{6}}\,(w_1 - w_2) p_3\,, 
\end{eqnarray} 
where, according to Eq.(\ref{vertex}), $\sum\limits_{i=1}^3w_ix_i=x$. Since 
the function
$\Phi\biggl(\sum\limits_{i<j}( x_i - x_j )^2 \biggr)$
is invariant under translations its Fourier transform only 
depends on two four--momenta. 
The function $\Phi(-l_1^2-l_2^2)$ in Eq.~(\ref{fourier}) 
will be modelled by a Gaussian form in our approach. 
The minus sign in the argument is chosen to emphasize
that we are working in Minkowski space.
Our choice is the Gaussian form 
\eq\label{corr_Fun}
\Phi(-l_1^2-l_2^2) = \exp\Big(18 (l_1^2+l_2^2)/\Lambda^2\Big) 
\en 
where the parameter $\Lambda$ characterizes the size of the DHB. 
Since $l_1^2$ and $l_2^2$ turn into $-l_1^2$ and $-l_2^2$ in 
Euclidean space the form~(\ref{corr_Fun}) has the appropriate fall--off
behavior in the Euclidean region. 

\subsection{Three--quark currents}

In the so--called $Q_1Q_2$--basis the DHBs  are classified by the set of 
quantum numbers $(J^P, S_d)$, where $J^P$ is the spin--parity of the baryon 
state and $S_d$ is the spin of the heavy diquark. There are two types of heavy 
diquarks -- those with $S_d = 0$ 
(antisymmetric spin  configuration $[Q_1Q_2]$) and those with $S_d = 1$ 
(symmetric spin 
configuration $\{Q_1Q_2\}$). Accordingly there are two $J^{P}=1/2\,^+$ 
DHB states. We follow the standard convention and attach 
a prime to the $S_d = 0$ states whereas the 
$S_d = 1$ states are unprimed. The 
$J^P = 3/2\,^+$ states are in the symmetric heavy quark spin 
configuration. 
In Table I we list the  
quantum numbers of the double--heavy baryons including their mass spectrum as 
calculated in~\cite{Ebert:2004ck}.

As we have discussed in our recent paper~\cite{Faessler:2009xn}, 
there is a mass inversion in the $(1/2^+)$ mixed flavor states
$(\Xi_{bc},\Xi'_{bc})$ and $(\Omega_{bc},\Omega'_{bc})$ in that
$M(\Xi'_{bc})>M(\Xi_{bc})$ and $M(\Omega'_{bc})>M(\Omega_{bc})$ even though
the heavy diquarks satisfy the conventional hyperfine splitting pattern 
$m_{(bc)_{S=1}}> m_{(bc)_{S=0}}$. 
This inversion is a feature of all models that have attempted to
calculate the mass spectrum of double--heavy baryons~\cite{Albertus:2006ya,
Ebert:2004ck,Kiselev:2001fw,Roberts:2007ni,Zhang:2008pm,Bernotas:2008fv}. 
In particular, the
inverted mass hierarchy implies that one can only expect substantial 
flavor--changing branching ratios for the two lowest lying states $\Xi_{bc}$ 
and $\Omega_{bc}$ whereas the rates of the higher lying states $\Xi_{bc}'$,
$\Xi_{bc}^{*}$ and $\Omega_{bc}'$, $\Omega_{bc}^{*}$ will be dominated
by flavor--preserving one-photon transitions to the lowest-lying states
$\Xi_{bc}$ and $\Omega_{bc}$. One of the purposes of the 
present paper is to analyze the strength of the one--photon
transitions between the $S_d = 0$ and $S_d = 1$ double--heavy baryon states.
In the HQL, the photon couples to the light quark only, and therefore 
one--photon transitions between the $S_d = 0$ and $S_d = 1$ double--heavy 
baryon states such as $\Xi'_{bc} \to \Xi_{bc}+\gamma$ are forbidden in this 
limit.
  For finite heavy quark masses one-photon transitions
between the $S_d = 0$ and $S_d = 1$ double--heavy baryon states will occur at 
a somewhat reduced rate which, however, is very likely to exceed the 
flavor--changing weak decay rates of these states \cite{Dai:2000hza}.   

Following the suggestion of Ref.~\cite{Roberts:2008wq} (see also 
discussion in~\cite{Albertus:2009ww}) we also consider hyperfine 
$\Xi_{bc}' - \Xi_{bc}$ and $\Omega_{bc}' - \Omega_{bc}$ mixing induced by
one--gluon exchange interactions.  
We define the mixed states through the unmixed states using a  
unitary transformation~\cite{Roberts:2008wq,Albertus:2009ww} with the
mixing angles $\theta_\Xi$ or $\theta_\Omega$:
\eq\label{hyp_mixing}
\left(
\begin{array}{c}
B_{bc}^h \\[3mm]
B_{bc}^l \\
\end{array}
\right) =
\left(
\begin{array}{cc}
  \cos\theta_B & \sin\theta_B \\[3mm]
- \sin\theta_B & \cos\theta_B \\
\end{array}
\right)
\left(
\begin{array}{c}
B_{bc}' \\[3mm]
B_{bc}  \\
\end{array}
\right)\,,
\en
where $B = \Xi$ or $\Omega$. We treat the mixing angle $\theta_B$ as
a quantity of order ${\cal O}(\alpha_s)$ where $\alpha_s$ is the
QCD coupling constant.
By $B^h$ and $B^l$ we denote the mixed states with the mass
hierarchy $m_{B_l} < m_{B} < m_{B'} < m_{B_h}$. The masses of the mixed
states $m_{B_h}$ and $m_{B_l}$ differ from the masses of the
unmixed states by small hyperfine splitting corrections. They
are expressed through the unmixed masses and mixing angle as:
\eq\label{masses_mixed}
m_{B_h} &=& m_{B'} + (m_{B'} - m_B) \frac{\sin^2\theta}{\cos2\theta} \,,
\nonumber\\
m_{B_l} &=& m_{B}  - (m_{B'} - m_B) \frac{\sin^2\theta}{\cos2\theta} \,.
\en
When diagonalizing the mass matrix of the unmixed states the diagonal elements
are driven apart such that
the mass difference of the mixed states is larger than that
of the unmixed states. One has 
\eq
m_{B_h} - m_{B_l} = \frac{m_{B'} - m_B}{\cos 2\theta} \,.
\en
This leads to a further enhancement of the widths of the radiative
transitions $B_{bc}^h \to B_{bc}^l + \gamma$ between mixed states,
because the photon transition rate
is proportional to $(M_{1}-M_{2})^{3}$ (see Appendix A).

The mixing angle $\theta_B$ corresponds to 
the combination $\theta + 30^0$ in~\cite{Albertus:2009ww}: 
$\theta_B \equiv \theta + 30^0$. Here $\theta$ is the angle  
that rotates the mixed $B^h, B^l$ states  
into the $\hat B, \hat B'$ states --- bound states 
of $b$-quark and heavy--light $cq$ diquark (so-called $qc$--basis). 
The angle $30^0$ corresponds to a further rotation of the $\hat B, \hat B'$ 
states into the unmixed $B', B$ states defined in the $bc$--basis 
(i.e. bound states of light quark and heavy $bc$ diquark).
In the quark model calculation of ~\cite{Albertus:2009ww}
one obtains  $\sin\theta_\Xi = 0.431$ ($\theta_\Xi = 25.5^0$) and 
$\sin\theta_\Omega = 0.437$ ($\theta_\Omega = 25.9^0$). 
Using these values of the mixing angle $\theta$ and the masses of unmixed
states we deduce the following values for the masses of the mixed states:
\eq
& &m_{\Xi_{bc}^h} = 6.972 \ {\rm MeV}\,, \quad\quad
   m_{\Xi_{bc}^l} = 6.924 \ {\rm MeV}\,, \nonumber\\
& &m_{\Omega_{bc}^h} = 7.125 \ {\rm MeV}\,, \quad\quad
   m_{\Omega_{bc}^l} = 7.079 \ {\rm MeV}\,.
\en

The interpolating currents of the DHB states 
$B_{qQ_1Q_2}$ are constructed
in the form of a light quark $q^{a_3}$ coupled to a heavy diquark 
$D^{a_3}_{Q_1Q_2}$. One obtains 
\eq 
J_{qQ_1Q_2} &=& \Gamma_{Q_1Q_2}^q q^{a_3}_3D_{Q_1Q_2}^{a_3}, \qquad
D_{Q_1Q_2}^{a_1}=\varepsilon^{a_1a_2a_3} \ \Big(Q_{1}^{a_1} 
C \Gamma^{D}_{Q_1Q_2} Q_2^{a_2}\Big) \,. 
\en  
For the $(\frac{1}{2}^+, 0)$, $(\frac{1}{2}^+, 1)$ and 
$(\frac{3}{2}^+, 1)$ states we use the simplest currents --- 
the pseudoscalar $J^P$, the vector $J^V$ and $J^V_\mu$ currents, 
respectively:  
\seq 
\eq 
J_{qQ_1Q_2}^P &=& \varepsilon^{a_1a_2a_3} \
q^{a_3} \Big(Q_{1}^{a_1} 
C \gamma_5 Q_2^{a_2}\Big) \,, \\
J_{qQ_1Q_2}^V &=& \varepsilon^{a_1a_2a_3} \ 
\gamma^\alpha\gamma^5  q^{a_3} \Big(Q_{1}^{a_1} 
C \gamma_\alpha Q_2^{a_2}\Big)\,,  \\
J_{qQ_1Q_2, \ \mu}^V &=& \varepsilon^{a_1a_2a_3} \ 
q^{a_3} \Big(Q_{1}^{a_1} 
C \gamma_\mu Q_2^{a_2}\Big)\,. 
\en  
\sen 
In the heavy quark limit the above currents reduce to 
\seq 
\eq 
j_{qQ_1Q_2}^P &=& \varepsilon^{a_1a_2a_3} \ 
\psi_{q}^{a_3} \ ( \psi_{Q_1}^{a_1} \, \sigma_2 \, \psi_{Q_2}^{a_2} ) \,, \\
j_{qQ_1Q_2}^V  &=& \varepsilon^{a_1a_2a_3} \ \vec{\sigma} 
\, \psi_{q}^{a_3} \ ( \psi_{Q_1}^{a_1} \, \sigma_2 \vec{\sigma} 
\, \psi_{Q_2}^{a_2} ) \,, \\ 
\vec{j}_{qQ_1Q_2}^{\, V} &=& \varepsilon^{a_1a_2a_3} \ \psi_{q}^{a_3} \ 
( \psi_{Q_1}^{a_1} \, \sigma_2 \vec{\sigma} \, \psi_{Q_2}^{a_2} ) \,, 
\en 
\sen 
where $\psi_{q,Q_1,Q_2}$ are the upper components of the Dirac quark spinors  
and the $\sigma_i$ are the Pauli spin matrices. Note that the spin--flavor 
wave function coincides with the nonrelativistic limit in the HQL. 
In the nonrelativistic limit our DHB currents have 
a one--to--one correspondence to the naive quark model baryon spin-flavor 
functions (up to overall factors) which are displayed in Table II. 
Further details on the naive quark model and how to evaluate the radiative 
transition amplitudes in this framework can be found in Appendix~B. 

According to the definition (\ref{hyp_mixing}) the interpolating currents of
the mixed $B^h$ and $B^l$ states are given by  
\eq\label{cur_mixing}
\left(
\begin{array}{c}
g_{B^h} J_{qbc}^{B^h} \\[3mm]
g_{B^l} J_{qbc}^{B^l} \\
\end{array}
\right) =
\left(
\begin{array}{cc}
  \cos\theta_B & \sin\theta_B \\[3mm]
- \sin\theta_B & \cos\theta_B \\
\end{array}
\right)
\left(
\begin{array}{c}
g_{B'} J_{qbc}^P  \\[3mm]
g_B    J_{qbc}^V  \\
\end{array}
\right)\,.
\en

\subsection{Baryon-quark coupling constant} 

The coupling constants 
$g_{\psi} \ (\psi = B, B^\ast)$ are determined by 
the compositeness condition~\cite{Faessler:2009xn,Ivanov:1996pz,%
Weinberg:1962hj,Efimov:1993ei}. The compositeness condition implies that the 
renormalization constant of the hadron wave function is set equal to zero,
i.e.  
\eq\label{ZLc}
Z_\psi = 1 - g_B^2 \Sigma_\psi^\prime(m_{\psi}) = 0 \,.
\en
$\Sigma_{\psi}^\prime(m_{\psi})$ 
is the derivative of the baryon mass operator shown in Fig.1. 

In case of the $3/2^+$ states the function $\Sigma_{B^\ast}(p)$ 
is subtracted from the nonvanishing part of the mass operator 
$\Sigma_{B^\ast}^{\mu\nu}(p)$ proportional to the Minkowski metric 
tensor~$g^{\mu\nu}$: 
\eq\label{Rarita_mass} 
\bar u_{\mu}(p,s^\ast) \Sigma_{B^\ast}^{\mu\nu}(p) 
u_{\nu}(p,s^\ast) 
= \bar u_{\mu}(p,s^\ast) \Sigma_{B^\ast}(p)g^{\mu\nu} 
u_{\nu}(p,s^\ast) \,,  
\en 
where $u^{\mu}(p,s^\ast)$ is the $\frac{3}{2}$--spinor.
Other possible Lorentz structure on the r.h.s. of Eq.~(\ref{Rarita_mass}) 
vanish due to the Rarita--Schwinger conditions. 
Note that the compositeness condition is equivalent to a Ward identity 
relating the electromagnetic vertex function at zero momentum transfer 
to the derivative of the mass operator (see details e.g. 
in~\cite{Faessler:2009xn}). Explicit expressions for the baryon mass 
operators are given in Appendix C.

\section{Radiative decays of double heavy baryons} 

\subsection{Matrix elements} 

In our approach the radiative decays of DHBs are described by the set of 
Feynman diagrams shown in Fig.2. The three ``triangle'' diagrams 
Figs.2(a)-2(c) are generated by the coupling of the constituent quarks with 
the photon. The two ``bubble'' diagrams in Figs.2(d) and 2(e) are
generated by gauging the nonlocal strong 
Lagrangian (see discussion in Sec.II). Finally the 
two ``pole'' diagrams in Figs.2(f) and 2(g) are generated by the direct 
coupling of the initial/final baryon with the photon. 
Due to our explicit construction the DHB 
radiative matrix elements are explicitly gauge-invariant. 
The pole diagrams vanish for the radiative transitions 
$(\frac{1}{2}^+, 0) \to (\frac{1}{2}^+, 1)$ and 
$(\frac{3}{2}^+, 1) \to (\frac{1}{2}^+, 0)$  
due to the orthogonality of the heavy diquark spin wave functions. 
For the same reason the photon does not couple to the light quarks in these
modes implying that the corresponding triangle and bubble graphs 
vanish. For the $(\frac{3}{2}^+, 1) \to (\frac{1}{2}^+, 1)$ transitions 
the pole diagram in Fig.2(g) vanishes due to the Rarita--Schwinger 
conditions for the $\frac{3}{2}$--spinor. All these statements are true
for photon transitions between unmixed states. Photon transitions between
mixed DHB states will be discussed in section IV.

We continue with a summary of some useful analytical results. 
The on--shell matrix elements for the radiative transitions
$\frac{1}{2} \to \frac{1}{2}$ and 
$\frac{3}{2} \to \frac{1}{2}$ are given by: 
\eq
M_{\mu}(1/2 \to 1/2) &=& 
\bar{u}_{B_2}(p_{2},s_{2}) \, \Lambda_{\mu}(p_{1},p_{2}) \, 
u_{B_1}(p_{1},s_{1})\,, 
\nonumber\\ 
M_{\mu}(3/2 \to 1/2) &=& 
\bar{u}_{B_2}(p_{2},s_{2}) \, \Lambda_{\mu\nu}(p_{1},p_{2}) \, 
u_{B_1^\ast}^\nu(p_{1},s^{\ast})\,, 
\en
where $u_B(p,s)$ and $u_{B^\ast}^\nu(p,s^\ast)$ are the spin $\frac{1}{2}$ 
and $\frac{3}{2}$ spinors with the normalization (see further details in 
Appendix~A): 
\eq 
\bar u_B(p,s) \, u_B(p,s) = 2 m_B\,, \quad\quad
\bar u_{B^\ast}^\mu(p,s^\ast) \, u_{B^\ast_\mu}(p,s^\ast) 
= - 2 m_{B^\ast} \,. 
\en 
The momenta of the final state photon, the initial and final state
baryon are denoted by $q$, $p_{1}$ and $p_{2}$, respectively, 
where $q = p_{1} - p_{2}$ and where $s$ and $s^\ast$ are spin indices. 
Due to gauge invariance the electromagnetic vertex function 
$\Lambda_{\mu}(p_{1},p_{2})$ is orthogonal to the photon
momentum $q^{\mu}\Lambda_{\mu}(p_{1},p_{2})=0$.
As a result the vertex function $\Lambda_\mu(p_{1},p_{2})$ is given
by the sum of the gauge-invariant pieces of the
triangle $(\Delta)$, the bubble (bub) and the pole (pol) diagrams, 
while the nongauge-invariant parts of these diagrams 
cancel in the sum: 
\eq
\Lambda_{\mu}(p_{1},p_{2}) =
\Lambda_{\mu, \, \Delta}^{\perp}(p_{1},p_{2}) +
\Lambda_{\mu, \, {\rm bub}}^{\perp}(p_{1},p_{2}) + 
\Lambda_{\mu, \, {\rm pol}}^{\perp}(p_{1},p_{2}) 
\en 
The contribution of each diagram can be split into gauge invariant and gauge 
variant pieces by introducing the decomposition 
\eq\label{split}
\gamma_\mu \, = \, \gamma_\mu^{\perp} \, + \,
q_\mu \, \frac{\not\! q}{q^2}\,,  \hspace*{1cm}
p_i \, = \, p_i^{\perp} \, + \, q_\mu \, \frac{p_{i} q}{q^2}\,, 
\hspace*{1cm} g_{\mu\nu} \, = \, g_{\mu\nu}^{\perp} \, + \, 
\frac{q_\mu \, q_\nu}{q^2} \,, 
\hspace*{1cm}
\en
such that $\gamma_\mu^\perp \,  q^\mu \, = \, 0$, 
$p_i^\perp \, q^\mu=0$, and 
$g_{\mu\nu}^{\perp} \, q^\mu=0$, where $p_{i}$ is  $p_{1}$ or $p_{2}$. 
The vertex function $\Lambda_\mu^\perp(p_{1},p_{2})$ can then be 
expressed in terms of $\gamma_\mu^\perp$, 
the $p_{i\,\mu}^\perp$ and $g_{\mu\nu}^{\perp}$. 
Note that all matrix elements are finite for real photons ($q^{2}=0$). 
Doing our calculations we start with $q^{2}\neq 0$ and then
take the limit $q^{2}\to 0$. 
Explicit expressions of the electromagnetic vertex functions 
can be found in Appendix~C.

\subsection{Heavy Quark Limit} 

In the HQL the masses of the heavy quarks are taken to infinity 
$(m_Q \to \infty)$. In this limit the spins of the double--heavy diquark and 
the light quark in the DHB states decouple leading to a much simplified  
transition structure. In particular, the transitions 
amplitudes $(\frac{1}{2}^+, 0) \to (\frac{1}{2}^+, 1)$ and 
$(\frac{3}{2}^+, 1) \to (\frac{1}{2}^+, 0)$ vanish as ${\cal O}(1/m_Q)$ 
since the photon coupling to the heavy quarks involves a spin--flip 
factor proportional to the magnetic moment of the heavy quark given by 
$\mu_Q = e_Q/(2m_Q)$. This is in full agreement with the 
nonrelativistic quark model (for more details see the discussion in 
Appendix B) where the corresponding amplitudes are found to be proportional 
to the difference of the magnetic moments of the heavy quarks 
$\mu_c - \mu_b$. The transition amplitude 
$(\frac{3}{2}^+, 1) \to (\frac{1}{2}^+, 1)$ survives in the HQL 
since the photon can now couple to the light quark. Again, 
this is consistent with the nonrelativistic quark model. The structure of 
the amplitude of the $(\frac{3}{2}^+, 1) \to (\frac{1}{2}^+, 1)$ transition 
significantly simplifies in the HQL. Only the 
triangle diagram in Fig.2(a) contributes to the transition amplitude since
Fig.2(a) represents the direct coupling of the light 
quark with the photon. The contribution of Fig.2(a) scales 
as ${\cal O}(1)$ in the inverse heavy quark mass expansion. 
The other diagrams are suppressed in the HQL. 
In particular, the triangle diagrams 
in Figs.2(b) and 2(c) contribute only at ${\cal O}(1/m_Q)$ since they
represent direct couplings of the heavy quarks 
with the photon. The same holds true for the bubble diagrams Figs.2(d) 
and 2(e) involving a nonlocal photon-light quark coupling. 

Following ideas developed in our paper~\cite{Faessler:2009xn}, 
we choose the momenta of the initial and final DHB as 
$p^\mu_1=(m_{Q_1}+m_{Q_2}) v^\mu$ 
and $p^\mu_2=(m_{Q_1} + m_{Q_2})v'^\mu=(m_{Q_1}+m_{Q_2}) v^\mu+r^\mu$ 
where $r$ is a small residual momentum in the sense that 
$r^2\sim {\cal O}(1)$ when $m_{Q_i}\to\infty$. 
with these assumptions the heavy quark propagators simplify 
in the HQL. One has
\eq
\tilde S_{Q_i}(k_i \pm p \eta_i) &\rightarrow&
\frac{1 \pm \not\! v}{2}\frac{1}{ \mp k_i v - i \epsilon},
\en
where $\eta_i = m_{Q_i}/(m_{Q_1}+m_{Q_2})$. 

In the HQL and at $q^2 = 0$ the explicitly gauge invariant transition 
amplitude $(\frac{3}{2}^+, 1) \to (\frac{1}{2}^+, 1)$ is given by 
\eq 
\Lambda_{\mu\nu}^\perp = \Gamma_{\mu\nu}^\perp \ I_\Delta\,, 
\hspace*{.5cm} I_{\Delta} = 
24 \, N_f \, e_q \, g_{B_1^\ast} \, g_{B_2} \, R_{12}(m_q,\Lambda) \,, 
\en 
where $N_f$ denotes a statistical flavor factor which is equal to 1 or 2 for 
DHBs with two different or two identical heavy quarks. 
$\Gamma_{\mu\nu}^\perp = (\gamma_\mu q_\nu - g_{\mu\nu} \!\! \not\! q) \, 
\gamma_5$ is the Lorentz structure orthogonal to the photon momentum: 
$q^\mu \, \Gamma_{\mu\nu}^\perp = 0$. Note that only the Lorentz structure 
$\Gamma_{\mu\nu}^\perp$ survives in the HQL. As shown in Appendix A other
possible structures vanish. The function $R_{12}(m_q,\Lambda)$ reads
\eq 
R_{AB}(m_q,\Lambda) = 
\int\frac{d^4k_1}{(2\pi)^4i} \, \int\frac{d^4k_2}{(2\pi)^4i} \, \Phi^2(z) \, 
\frac{m_q+(k_2-k_1)v }{(- k_1v - i\epsilon)^A (k_2v - i\epsilon)
(m_q^2 - (k_2 - k_1)^2 - i\epsilon)^B} 
\en 
where $z = - \frac{2}{3} (k_1^2 - k_1 k_2 + k_2^2)$ and where $A$ and $B$
denote integer powers. 
The coupling constants $g_{B_1^\ast}$ and $g_{B_2}$ are given by 
\eq 
\frac{1}{g_{B_1^\ast}^2} =  
\frac{1}{3 g_{B_2}^2} = 12 \, N_f \, \, R_{21}(m_q,\Lambda) \,. 
\en 
Using the Laplace transform 
\eq
\Phi^2(z) = \int\limits_0^\infty ds \, \Phi^L(s) \, e^{-sz}
\en
the integration over the virtual momenta $k_1$ and $k_2$ 
in $R_{AB}(m_q,\Lambda)$ can be done. One obtains 
\eq 
R_{21}(m_q,\Lambda) 
&=& \frac{\Lambda^4}{(16 \pi^2)^2}
 \, \int\limits_0^\infty 
d \alpha_1 d \alpha_2 d \alpha_3 \frac{\alpha_1}{D^2} \, 
\biggl( \frac{m_q}{\Lambda} 
+ \frac{\alpha_1 + \alpha_2}{4 D} \biggr)  \, \Phi^2(y) \,, \nonumber\\
R_{12}(m_q,\Lambda) &=& \frac{\Lambda^3}{(16 \pi^2)^2} \, 
\int\limits_0^\infty 
d \alpha_1 d \alpha_2 d \alpha_3 \frac{\alpha_3}{D^2} \, 
\biggl( \frac{m_q}{\Lambda} 
+ \frac{\alpha_1 + \alpha_2}{4 D} \biggr)  \, \Phi^2(y) 
\,, 
\en 
where 
\eq 
D = \frac{3}{4} + \alpha_3\,, \hspace*{.5cm} 
\frac{y}{\Lambda^2} = 
\frac{2}{3} \, \biggl( \frac{m_q^2}{\Lambda^2} \, \alpha_3 
+ \frac{(1+\alpha_3) (\alpha_1 - \alpha_2)^2 + \alpha_1\alpha_2}{4 D}
\biggr) \, . 
\en  
For the $(\frac{3}{2}^+, 1) \to (\frac{1}{2}^+, 1)$ transition the HQL
helicity amplitudes read   
\eq 
\label{hel1}
H_{\pm \frac{1}{2} \mp 1} &=& 
\bar\epsilon^{\,\ast\mu}(\mp 1) 
\bar u\Big(v, \pm \frac{1}{2}\Big) 
\Lambda_{\mu\nu}^\perp \, 
u^\nu\Big(v, \pm \frac{3}{2}\Big) 
= \pm M_+ M_- \ \sqrt{\frac{M_2}{M_1}} \ I_\Delta \,,  \\
\label{hel2}
H_{\pm \frac{1}{2} \pm 1} &=& 
\bar\epsilon^{\,\ast\mu}(\pm 1) 
\bar u\Big(v, \pm \frac{1}{2}\Big) 
\Lambda_{\mu\nu}^\perp \, 
u^\nu\Big(v, \mp \frac{1}{2}\Big) 
= \mp \sqrt{\frac{1}{3}} \, M_+ M_- \ \sqrt{\frac{M_2}{M_1}} \ I_\Delta \,. 
\en 
The ratio of the HQL helicity amplitudes is given by 
$H_{\pm \frac{1}{2} \mp 1}/H_{\pm \frac{1}{2} \pm 1} = - \sqrt{3}$, i.e. 
one has a pure $M1$ magnetic dipole transition. This 
coincides with the predictions of the nonrelativistic quark model (NQM) 
(see Appendix B), where  
\eq 
H_{\pm \frac{1}{2} \mp 1} = - \sqrt{3} \, H_{\pm \frac{1}{2} \pm 1} 
= \pm \frac{2 \mu_q}{\sqrt{3}} \, 
M_+ M_- \ \sqrt{\frac{M_2}{M_1}} \,. 
\en 
The function $I_\Delta$ appearing in (\ref{hel1}) and (\ref{hel2}) 
is given by 
\eq\label{beta_factor} 
I_\Delta = \frac{2 \mu_q}{\sqrt{3}} \beta \,, 
\en  
where 
\eq\label{beta_par}
\beta = 2 m_q \frac{R_{12}(m_q,\Lambda)}{R_{21}(m_q,\Lambda)} \,. 
\en 
In the HQL the helicity amplitudes, the form factors $F_1$ 
and $F_2$, and the decay rate 
of the $(\frac{3}{2}^+, 1) \to (\frac{1}{2}^+, 1)$ transition read 
\eq 
& &H_{\pm \frac{1}{2} \mp 1} = - \sqrt{3} \, H_{\pm \frac{1}{2} \pm 1}  
= \pm \frac{2 \mu_q}{\sqrt{3}} \, \beta \, 
M_+ M_- \ \sqrt{\frac{M_2}{M_1}}  \,, \nonumber\\ 
& &F_1 = \frac{2 \mu_q}{\sqrt{3}}\, \beta \, M_+ \ \sqrt{\frac{M_2}{M_1}}\,, 
\quad\quad 
F_2 = - \frac{4 \mu_q}{\sqrt{3}}\, \beta \, \frac{M_2}{M_+} 
\ \sqrt{M_1M_2} \,, \\ 
& &\Gamma_{\frac{3}{2} \to \frac{1}{2}^S} = \frac{4}{3} \, 
K \, \mu_q^2 \, \beta^2 \,, \nonumber
\en  
where 
\eq 
K = \alpha \, M_2 \, \frac{(M_1^2-M_2^2)^3}{6 M_1^4}  \,. 
\en 
$\alpha \simeq  1/137$ is the fine structure coupling constant and 
$M_1$, $M_2$ are the masses of the parent and daughter 
baryon. 
The state $\frac{1}{2}^S$ corresponds to baryon with symmetric 
heavy quark spin configuration. 

It is evident that our HQL rate results differ from the predictions 
of the NQM by the factor $\beta^2$ defined in (\ref{beta_par}). In the NQM 
one has $\beta \equiv 1$ while in our covariant 
approach $\beta \approx 0.5$. It is for this reason that 
our HQL predictions for the $(\frac{3}{2}^+, 1) \to (\frac{1}{2}^+, 1)+\gamma$ 
decay widths are down by a factor of 4 compared to the 
predictions of the NQM. 

\section{Hyperfine mixing and radiative decays of mixed states}
\label{hyp_mix} 
As mentioned in the introduction the origin of the hyperfine mixing in the 
double heavy baryons is the one--gluon exchange interaction between the light 
and heavy quarks in the states containing two different heavy quarks --- $b$ 
and $c$. The one--gluon interaction leads to
mixing of the states containing spin--0 and spin--1 heavy
quark configurations.
In this section we discuss in some detail the calculations of 
DHB radiative decays involving mixed states. We have three types of
transitions: $B^h_{bc} \to B^l_{bc}$, $B^\ast_{bc} \to B^l_{bc}$ 
and $B^\ast_{bc} \to B^h_{bc}$. All three modes are quite interesting, 
because their study opens the opportunity to determine the mixing 
angle $\theta$ and to measure the masses of the mixed states. 
In particular, the first mode  $B^h_{bc} \to B^l_{bc}$ is interesting since
it is described by transitions between baryon components with the same spin 
configuration of the heavy quarks $(\frac{1}{2}^+, 1) \to (\frac{1}{2}^+, 1)$ 
and $(\frac{1}{2}^+, 0) \to (\frac{1}{2}^+, 0)$. Because now the photon
can also couple to the light quark one will have a corresponding enhancement
of the decay rates. 
The two other modes $B^\ast_{bc} \to B^l_{bc}$ 
and $B^\ast_{bc} \to B^h_{bc}$ involve mixing of the leading 
$(\frac{3}{2}^+, 1) \to (\frac{1}{2}^+, 1)$ 
and subleading $(\frac{3}{2}^+, 1) \to (\frac{1}{2}^+, 0)$  
amplitudes and are therefore also important for an analysis of 
the mixing angle $\theta$. 
The matrix elements for transitions involving mixed states are 
derived using the transition matrix elements of the unmixed 
states. In Appendix B we present the results of the NQM for  
transitions involving mixed states in terms of quark magnetic moments 
and the mixing angle $\theta$. In our numerical calculations we differentiate 
between the mixing angles for the $\Xi$--states and the $\Omega$--states using 
the predictions of the quark model~\cite{Albertus:2009ww}: 
$\theta_\Xi \simeq 25.5^0$ and $\theta_\Omega \simeq 25.9^0$.  

The last issue which we would like to discuss in this section is 
the HQL structure of transitions involving mixed states. As we have discussed
in the previous 
section the leading contribution for the $B^\ast_{bc} \to B^l_{bc}$ and 
$B^\ast_{bc} \to B^h_{bc}$ transitions comes from the 
$(\frac{3}{2}^+, 1) \to (\frac{1}{2}^+, 1)$ transition generated 
by the direct coupling of the photon with the light quark 
(see diagram in Fig.2(a)). The corresponding amplitudes  
are multiplied by the factor $\cos\theta$ for the 
$B^\ast_{bc} \to B^l_{bc}$ mode and by $-\sin\theta$ for the 
$B^\ast_{bc} \to B^h_{bc}$ mode. In the case of the
$B^h_{bc} \to B^l_{bc}$ transition the leading contribution 
is again generated by the direct light quark--photon coupling 
[Fig.2(a)]. In this case one has to sum the two transitions 
involving a light quark spin--flip  
$(\frac{1}{2}^+, 0) \to (\frac{1}{2}^+, 0)$ and 
$(\frac{1}{2}^+, 1) \to (\frac{1}{2}^+, 1)$. The calculation of these 
leading matrix elements follows the treatment in the previous 
section. In particular, the leading contribution to the matrix element of the 
$B^h_{bc} \to B^l_{bc}$ transition is expressed though the same 
structure integral $R_{12}(m_q,\Lambda)$ as in the case of the
$(\frac{3}{2}^+, 1) \to (\frac{1}{2}^+, 1)$ transition 
\eq 
\Lambda_\mu^\perp = \not\! q \gamma_\mu^\perp \, J_\Delta\,, 
\hspace*{.5cm} J_{\Delta} = 
24 \, N_f \, e_q \, \sin(2\theta_B) \, \frac{g_{B}^2 + g_{B'}^2}{2} 
\, R_{12}(m_q,\Lambda) \,, 
\en 
where $g_B$ and $g_B'$ are the coupling constants of the unmixed states 
$(\frac{1}{2}^+, 1)$ and $(\frac{1}{2}^+, 0)$: 
\eq 
\frac{1}{g_{B'}^2} =  
\frac{1}{3 g_{B}^2} = 12 \, N_f \, \, R_{21}(m_q,\Lambda) \,. 
\en  
In the HQL the helicity amplitudes of the $B^h_{bc} \to B^l_{bc}$ transition 
are given by 
\eq 
H_{\pm \frac{1}{2} \pm 1} = 
\bar\epsilon^{\,\ast\mu}(\pm 1) 
\bar u\Big(v, \pm \frac{1}{2}\Big) 
\Lambda_\mu^\perp \, 
u\Big(v, \mp \frac{1}{2}\Big) 
= M_+ M_- \ \sqrt{\frac{2M_2}{M_1}} \ J_\Delta \,. 
\en 
After some straightforward algebra one can express the helicity amplitudes
$H_{\pm \frac{1}{2} \pm 1}$ 
in terms of the parameter $\beta$ derived in Eq.~(\ref{beta_par}). One has 
\eq 
H_{\pm \frac{1}{2} \pm 1} 
= \frac{2\sqrt{2}}{3} \, \mu_q \, \sin\! 2\theta \, \beta \, 
 M_+ M_- \ \sqrt{\frac{M_2}{M_1}} \,. 
\en 
For $\beta \equiv 1$ our helicity amplitudes coincide with the predictions 
of the NQM. We can also deduce the form factor $F_2$ [see the expression 
for the matrix element of the $1/2^+ \to 1/2^+$ transition~(\ref{12_12})]: 
\eq 
F_2 =  - \frac{2}{3} \, \mu_q \, \sin\! 2\theta \, \beta \, M_1 \, 
\sqrt{\frac{M_2}{M_1}} \,. 
\en  
Note that the form factor $F_1$ defined in (\ref{12_12}) vanishes  
due to gauge invariance. 
Finally the decay width for the $B^h_{bc} \to B^l_{bc}$ 
transition in the HQL reads
\eq 
\Gamma(B^h_{bc} \to B^l_{bc}) = \frac{4}{3} 
\, K \, \mu_q^2 \sin^2\!2\theta \,,
\en 
which again coincides with the prediction of the NQM 
[see Eq.~(\ref{rates_NQM})] when $\beta \equiv 1$.   

Note that in the HQL our model also reproduces the model--independent 
results derived in~\cite{Albertus:2009ww} 
for the decay rates involving mixed states $\frac{1}{2}^{h}$ 
and $\frac{1}{2}^{l}$: 
\eq 
\Gamma_{\frac{1}{2}^{h} \to \frac{1}{2}^{l}} &\sim& \mu_q^2 \, 
\sin^2\!2\theta \,, \nonumber\\
\Gamma_{\frac{3}{2} \to \frac{1}{2}^{l}} &\sim& \mu_q^2 \, 
\cos^2\!\theta \,, \nonumber\\
\Gamma_{\frac{3}{2} \to \frac{1}{2}^{h}} &\sim& \mu_q^2 \, 
\sin^2\!\theta \,, \nonumber\\
\displaystyle{\frac{\Gamma_{\frac{3}{2} \to \frac{1}{2}^{h}}} 
                   {\Gamma_{\frac{3}{2} \to \frac{1}{2}^{l}}}} 
&\sim& \tan^ 2\!\theta \,. 
\en 
 
\section{Results}

We now proceed to present our numerical results. We first present results
on the radiative rates using finite heavy quark masses, i.e. we
do not take the HQL for the matrix elements. Estimates for the decay widths 
are also given for the nonrelativistic quark model,
in which, as described before, the wave functions have the same spin--flavor
structure as our relativistic current considered in the nonrelativistic limit. 
Then we consider the HQL in both approaches. 
We choose the Gaussian form Eq.~(\ref{corr_Fun}) for the correlation function
of the double heavy baryons. Our results depend on
the following set of parameters: the constituent quark masses and
the size parameter $\Lambda_B$. The parameters have been taken from a fit to
the properties of light, single and double heavy
baryons in previous analyses~\cite{Ivanov:1996pz}:
\begin{equation}
\begin{array}{cccccc}
m_{u(d)} & m_s & m_c & m_b & \Lambda_B \\  
$\ \ 0.42\ \ $ & $\ \ 0.57\ \ $ & $\ \ 1.7\ \ $ & $\ \ 5.2\ \ $
& $\ \ 2.5 \ $ - $ \ 3.5\ \ $ & $\ \ {\rm GeV} $ \\
\end{array} \ \ \,.
\label{sizepar}
\end{equation}
All our analytical calculations have been done using the computer program
FORM~\cite{Vermaseren:2000nd}. 

In Table III we present detailed numerical results on the
radiative rates of double heavy baryons  
using finite masses for the heavy quarks (exact results, second column) 
and in the HQL (third column). These are compared to the corresponding 
results of the NQM using finite masses for the heavy quarks (fourth column).
In column 5 we take the heavy quark limit of the NQM results by setting 
the heavy quark magnetic moments to zero. The dependence of our results on 
the size parameter $\Lambda_{B}$ is indicated by error bars where the 
variation of $\Lambda_{B}$ is given in Eq.(\ref{sizepar}) . Note that a smaller
value of $\Lambda$ gives bigger rates and vice versa. One can see that our 
finite heavy quark mass predictions are close to the results 
of the NQM. 
One has to keep in mind that the rate predictions are very sensitive to
the mass difference $\Delta M=M_{1}-M_{2}$. In fact, one has 
$\Gamma \sim (\Delta M)^{3}$ (see Appendix B). 
The modes involving mixed states are 
enhanced by factors $\simeq 4$ and $\simeq 2$ in case of
$B_{bc}^h \to B_{bc}^l$ and $B_{bc}^\ast \to B_{bc}^l$ transitions,
respectively, while in case of $B_{bc}^\ast \to B_{bc}^h$ transitions
they are additionally suppressed by a factor $\simeq 10$ due
to reduction of the mass difference $M_1 - M_2$. 
In Table IV we present results for the rates involving mixed states
in dependence on the mixing angle $\theta_B$ varied from $10^0$ to $25^0$. 
In Table V we compare our results with the results  
of a nonrelativistic quark model calculation~\cite{Albertus:2010hi}. The 
approach ~\cite{Albertus:2010hi} can be viewed 
as an extension of the naive NQM discussed before by taking into account 
baryon wave functions in configuration space. 
One should emphasize that the results of nonrelativistic quark models are
in general frame-dependent. For example, the results can
depend on whether one works in the parent baryon or daughter baryon rest
frame. Also, it is difficult to maintain gauge invariance in nonrelativistic 
quark models.

One final remark concerns the comparison of radiative and 
weak decays of DHBs. In~\cite{Faessler:2009xn}  
we have calculated the $b \to c$ semileptonic decays where we have shown that 
the corresponding decay widths are of the order of $10^{-14}$ GeV. 
The radiative decay widths calculated in the present paper are much larger and 
lie in the range from $10^{-8}$ to $10^{-4}$ GeV. One would like to know
how important the weak decays of DHBs induced by the 
$q_i \to q_j$ ($d \to u$ 
or $s \to u$) light quark transitions are. For a precise analysis 
one would need to know the precise values of the masses of the DHBs 
including isospin--breaking corrections which are not available at present.
Instead using the general formula for the semileptonic decay width  
one can obtain a rough estimate for the decay rates induced by 
light quark transitions where, for the sake of simplicity, we neglect the 
contribution of form factors, spin and flavor factors. For example, for the 
$1/2^{+} \to 1/2^{+}$
transition one obtains (see e.g. \cite{Kadeer:2005aq})
\eq
\Gamma(q_i \to q_j) 
\cong  \frac{G_F^2 |V_{\rm CKM}|^2}{15 \pi^3} \Delta M^5 \,, 
\en
where $G_F = 1.16634 \times 10^{-5}$ GeV$^{-2}$ is the Fermi constant, 
$V_{\rm CKM}$ is Cabibbo--Kobayashi--Maskawa (CKM) matrix element 
($|V_{ud}|^2 \simeq 1$ and $|V_{us}|^2 = 0.051$), and $\Delta M = M_1 - M_2$ 
is the difference of the masses of initial and final baryons. 
We know from data on the mass differences of light and heavy--light baryons 
that, approximately, 
the mass difference 
$\Delta M$ does not exceed the mass difference of the 
corresponding light quarks in these baryons. Therefore, in the expression for 
$\Gamma(q_i \to q_j)$ we substitute $\Delta M \leq m_d - m_u$ 
for $d \to u$ transitions and $\Delta M \leq m_s - m_u$ 
for $s \to u$ transitions and considering 
this to be an approximation of the upper limit for the 
corresponding decay rates. Using upper limits $m_d - m_u < 10$~MeV 
and $m_s - m_u < 200$~MeV we obtain: 
\eq
\Gamma(d \to u) < 10^{-22} \ {\rm GeV}\,, \hspace*{1cm}
\Gamma(s \to u) < 10^{-18} \ {\rm GeV}\,. 
\en
Based on this rough estimate one concludes that the semileptonic decays of 
DHBs induced by the light 
quark transitions $d \to u$ and $s \to u$ are suppressed 
by more than 4 orders of magnitude in comparison to their $b \to c$ 
counterparts and are even more suppressed (more than 10 orders) 
compared to their radiative decays. 

\section{Summary}

We have analyzed the radiative decays of double heavy baryons using a 
manifestly Lorentz covariant and gauge invariant constituent quark model 
approach. Our main results can be summarized as follows. 
We have derived results for the radiative transition matrix elements of
double heavy baryons for finite values of the heavy quark/baryon masses and 
also in the HQL limit of infinitely heavy quark masses. We have discussed
in detail radiative transitions involving DHB states subject to hyperfine
mixing. We have presented an extensive numerical analysis of the decay rates
for finite masses and in the HQL limit including numerical results on mixing
effects. Our results were compared with the predictions of a 
nonrelativistic quark model including again hyperfine mixing effects.  
We find that the inclusion of hyperfine mixing effects has a profound  
influence on the pattern of radiative decays of DHBs. Since the calculated
rates depend very sensitively on the exact mass values of the mixed and 
unmixed DHB states ($\Gamma \sim (M_{1}-M_{2})^{3}$) one must wait for an 
accurate determination of the masses of the DHB states before one can 
extract information on the mixing angles from the decay data.  

\begin{acknowledgments}

This work was supported by the DFG under Contract No. FA67/31-2
and No. GRK683. M.A.I. appreciates the partial support by
the DFG grant KO 1069/13-1, the Heisenberg-Landau program
and the Russian Fund of Basic Research grant No. 10-02-00368-a.
This research is also part of the European
Community-Research Infrastructure Integrating Activity
``Study of Strongly Interacting Matter'' (acronym HadronPhysics2,
Grant Agreement No. 227431), Russian President grant
``Scientific Schools''  No. 3400.2010.2, Russian Science and
Innovations Federal Agency contract No. 02.740.11.0238.

\end{acknowledgments}

\appendix 
\section{Spin--kinematics of radiative decays} 

In this Appendix we write down covariant expansions for the 
current--induced electromagnetic transitions involving the $(1/2^{+})$ 
and $(3/2^{+})$ baryon states. We thereby define sets of invariant vector 
transition form factors. We then define helicity amplitudes which are 
expressed in terms of linear combinations of the invariant form 
factors. One of the advantages of using helicity amplitudes is that one 
obtains very compact expressions for the decay rates 
(see e.g. \cite{Kadeer:2005aq,Bialas:1992ny}). In addition, the helicity
amplitudes contain the complete spin information of the process and are thus
well suited for the computation of spin observables.

In the radiative decays of double--heavy 
baryons the momenta and masses are denoted by 
\eq
B_1(p_1,M_1) \to B_2(p_2,M_2) + \gamma(q) \, 
\en 
where $p_1=p_2+q$. 
For the invariant form factor expansion of the $1/2^{+} \to 1/2^{+}$ matrix 
elements of the vector current $J_\mu$ one obtains

Transition $\frac{1}{2}^+ \to \frac{1}{2}^+$\,: 
\eq\label{12_12} 
M_\mu = \la B_2 | J_\mu | B_1 \ra = \bar u(p_2,s_2) 
\Big[ \gamma_\mu F_1(q^2) 
    - i \sigma_{\mu\nu} \frac{q_\nu}{M_1} F_2(q^2)  
    + \frac{q_\mu}{M_1} F_3(q^2) 
\Big] u(p_1,s_1) \,.
\en 

Similarly one has 

Transition $\frac{3}{2}^+ \to \frac{1}{2}^+$\,: 
\eq\label{32_12}  
M_\mu = 
\la B_2 | J_\mu | B_1^\ast \ra = \bar u(p_2,s_2) 
\Big[                      g_{\alpha\mu} F_1(q^2)  
    + \gamma_\mu \frac{p_{2\alpha}}{M_2} F_2(q^2) 
+ \frac{p_{2\alpha} p_{1\mu}}{M_2^2}     F_3(q^2)   
    + \frac{p_{2\alpha} q_\mu}{M_2^2}    F_4(q^2)       
\Big] \gamma_5 u^\alpha(p_1,s_1) \,,
\en 
where 
$\sigma_{\mu\nu} = (i/2) (\gamma_\mu \gamma_\nu - \gamma_\nu \gamma_\mu)$ 
and all $\gamma$ matrices are defined as in Bjorken--Drell. One should 
emphasize that the above invariant form factors are constrained by gauge 
invariance relations 
(see e.g. the detailed discussion in~\cite{Devenish:1975jd}). 

Next we express the vector helicity amplitudes 
$H_{\lambda_2\lambda_\gamma}$ in terms of the invariant form factors 
$F_i$, 
where $\lambda_\gamma = \pm 1$ and $\lambda_2 = \pm 1/2, \pm 3/2$ are 
the helicity components of the on--shell photon and the daughter baryon, 
respectively. The pertinent relation is
\eq 
H_{\lambda_2\lambda_\gamma} = M_\mu(\lambda_2) 
\bar\epsilon^{\,\ast\mu}(\lambda_\gamma) \,. 
\en 
Angular momentum conservation fixes the helicity $\lambda_1$ of the parent 
baryon according to $\lambda_1 = \lambda_2 - \lambda_\gamma$. We shall work 
in the rest frame of 
the parent baryon $B_1$ with the daughter baryon $B_2$ moving in the 
positive $z$-direction such that
$p_1^\mu = (M_1, {\bf 0})$, $p_2^\mu = (E_2, 0, 0, |{\bf p}_2|)$ and 
$q^\mu = (q_0, 0, 0, - |{\bf p}_2|)$, 
where $q_0 = |{\bf p}_2| = (M_1^2 - M_2^2)/(2 M_1)$ and 
$E_2 = M_1 - q_0 = (M_1^2 + M_2^2)/(2 M_1)$. 

The $J=\frac{1}{2}$ baryon spinors are given by 
\eq 
\bar u_2\Big(p_2, \pm \frac{1}{2}\Big) &=& \sqrt{E_2 + M_2} 
\Big( \chi_\pm^\dagger, \frac{\mp |{\bf p}_2|}{E_2 + M_2}  
\chi_\pm^\dagger \Big)\,, \nonumber\\
u_1\Big(p_1, \pm \frac{1}{2}\Big) &=& \sqrt{2M_1} 
\left(
\begin{array}{l}
\chi_\pm \\
0 \\
\end{array}
\right)
\en 
where $\chi_+ = \left(
\begin{array}{l}
1 \\
0 \\
\end{array} \right)$ 
and $\chi_- = \left(
\begin{array}{l}
0 \\
1 \\
\end{array} \right)$ are two--component Pauli spinors. 

The $J=\frac{3}{2}$ baryon spinors are 
defined by 
\eq 
u_\mu(p,s^\ast) = \sum\limits_{\lambda,s} 
\la 1\lambda\frac{1}{2}s | \frac{3}{2} s^\ast \ra 
\epsilon_\mu(p,\lambda) u(p,s)\,. 
\en  
They satisfy the Rarita--Schwinger conditions 
\eq\label{Rarita_Schwinger} 
\gamma^\mu \, u_\mu(p,s^\ast) = p^\mu \, u_\mu(p,s^\ast) = 0 \,, 
\en
where $\la 1\lambda\frac{1}{2}s | \frac{3}{2} s^\ast \ra$ is 
the requisite Clebsch-Gordan coefficient, $\epsilon_\mu(p,\lambda)$ is the 
spin 1 polarization 
vector and $u(p,s)$ are the usual $J=\frac{1}{2}$ spinors  
defined above. In particular, the $J=\frac{3}{2}$ spinors with 
helicities $\lambda=\pm3/2,\pm1/2$ read: 
\eq 
& &u_\mu\Big(p, \pm \frac{3}{2}\Big) 
= \epsilon_\mu(p, \pm 1) u\Big(p, \pm \frac{1}{2}\Big) \,, 
\nonumber\\
& &u_\mu(p, \pm \frac{1}{2}) 
= \sqrt{\frac{2}{3}} \epsilon_\mu(p, 0) 
u\Big(p, \pm \frac{1}{2}\Big) 
+ \sqrt{\frac{1}{3}} \epsilon_\mu(p, \pm 1) 
u\Big(p, \mp \frac{1}{2}\Big) \, . 
\en
The polarization vectors corresponding to the parent and daughter 
$J=\frac{3}{2}$ baryons are given by: 
\eq 
& &\,\,\epsilon^\mu(p_1,0) = (0,0,0,1)\,, \hspace*{2.05cm}  
\epsilon^\mu(p_1,\pm 1) = \frac{1}{\sqrt{2}} (0,\mp 1,-i,0)\,, \nonumber\\
& &\epsilon^{\ast\mu}(p_2,0) = \frac{1}{M_2} (|{\bf p}_2|,0,0,E_2)\,, 
\hspace*{.5cm}   
\epsilon^{\ast\mu}(p_2,\pm 1) = \frac{1}{\sqrt{2}} (0,\mp 1,i,0)  \,. 
\en 
The polarization vectors of the on--shell photon read 
\eq 
\bar\epsilon^{\,\ast\mu}(\pm 1) &=& \frac{1}{\sqrt{2}}(0, \pm 1, i, 0) \,. 
\en 
where the ``bar'' on the polarization vector denotes the fact that the photon 
is moving in the negative $z$--direction.
The polarization vectors satisfy the Lorentz condition  
\eq 
q_\mu \bar\epsilon^{\,\ast\mu}(\pm 1) = 0\,. 
\en 
Using above formulas for the spin wave functions with definite helicities
one can then express the helicity amplitudes 
$H_{\lambda_2\lambda_\gamma}$ through the invariant form factors by
calculating
$H_{\lambda_2\lambda_\gamma} = M_\mu(\lambda_2) 
\bar\epsilon^{\,\ast\mu}(\lambda_\gamma)$. One obtains 

Transition $\frac{1}{2}^+ \to \frac{1}{2}^+$\,:
\eq 
H_{\pm \frac{1}{2}\pm 1} = - F_2 \sqrt{2} \, \frac{M_+ M_-}{M_1} \,, 
\en 
where $M_\pm = M_1 \pm M_2$. 

Transition $\frac{3}{2}^+ \to \frac{1}{2}^+$\,:
\eq\label{helicity_31}
H_{\pm \frac{1}{2} \pm 1} &=& \pm \, 
\frac{1}{\sqrt{3}} \, M_- \, 
\Big( F_1 + F_2 \, \frac{M_+^2}{M_1 M_2}  \Big) \,,\nonumber\\
H_{\pm \frac{1}{2} \mp 1} &=& \pm \, M_- \, F_1 \,. 
\en
Note that one has an explicit factor of $M_{-}$ in all the helicity
amplitudes which corresponds to the appropriate $p$--wave threshold factor
$| {\bf p} | \sim M_{-}$. 
The decay width is given by  
\eq 
\Gamma_{s_1 \to s_2} = \frac{\alpha}{2s_{1}+1} \ 
\frac{M_1^2 - M_2^2}{4 M_1^3}  \ {\cal H}_{s_1 \to s_2}     
\en 
where the 
${\cal H}_{s_1 \to s_2}$ are bilinear combinations of 
the helicity amplitudes: 
\eq 
{\cal H}_{\frac{1}{2} \to \frac{1}{2}} &=&  
  |H_{\frac{1}{2}1}|^2 +  |H_{-\frac{1}{2}-1}|^2 \,, \nonumber\\
{\cal H}_{\frac{3}{2} \to \frac{1}{2}} &=&  
|H_{\frac{1}{2}1}|^2  + |H_{-\frac{1}{2}-1}|^2 +  
|H_{\frac{1}{2}-1}|^2 + |H_{-\frac{1}{2}1}|^2  \,. 
\en 
The overall dependence of the rate on the mass difference $M_{-}$ can be seen
to be given by $| {\bf p} |^{\,2l+1} \sim M_{-}^{3}$ for $l=1$. 
In the HQL the baryon spinors simplify. For example, 
the HQL $J=\frac{1}{2}$ baryon spinors read
\eq 
\bar u_2\Big(v, \pm \frac{1}{2}\Big) = \sqrt{2M_2} 
\Big( \chi_\pm^\dagger, 0 \Big)\,, \hspace*{.5cm}
u_1\Big(v, \pm \frac{1}{2}\Big) = \sqrt{2M_1} 
\left(
\begin{array}{l}
\chi_\pm \\
0 \\
\end{array}
\right) \,. 
\en 
For the $\frac{3}{2}^+ \to \frac{1}{2}^+$ transition in the HQL 
one has four possible Dirac strings in the matrix elements which are 
$\gamma_5$, $\gamma_5 \! \not\! q$, 
$\gamma_5 \! \not\!\bar\epsilon^{\,\ast}(\lambda_\gamma)$ and 
$\gamma_5 \! \not\! q \not\!\bar\epsilon^{\,\ast}(\lambda_\gamma)$. 
It is easy to show that the two Dirac strings $\gamma_5$ and 
$\gamma_5 \! \not\! q \not\!\bar\epsilon^{\,\ast}(\lambda_\gamma)$
vanish when sandwiched between the HQL baryon spinors: 
\eq 
\bar u_2\Big(v, \pm \frac{1}{2}\Big) \Big(\gamma_5,
\gamma_5 \! \not\! q \not\!\bar\epsilon^{\,\ast}(\lambda_\gamma)\,\Big) 
u_\mu\Big(v, \pm \frac{3}{2}\Big) = 
\bar u_2\Big(v, \pm \frac{1}{2}\Big)\Big(\gamma_5,
\gamma_5 \! \not\! q \not\!\bar\epsilon^{\,\ast}(\lambda_\gamma)\,\Big)  
u_\mu\Big(v, \mp \frac{1}{2}\Big) = 0 \,. 
\en 
The remaining strings $\gamma_5 \not\! q$ and 
$\gamma_5 \not\!\bar\epsilon^{\,\ast}$ 
one has
\eq 
& &\bar u_2\Big(v, \pm \frac{1}{2}\Big)
\gamma_5 \not\! q \, 
u_\mu\Big(v, \pm \frac{3}{2}\Big) = 
\mp M_+ M_- \ \sqrt{\frac{M_2}{M_1}} \, 
\epsilon_\mu(v, \pm 1)\,, \nonumber\\ 
& &\bar u_2\Big(v, \pm \frac{1}{2}\Big) 
\gamma_5 \not\! q \, 
u_\mu\Big(v, \mp \frac{1}{2}\Big) = \mp \sqrt{\frac{1}{3}} 
M_+ M_- \ \sqrt{\frac{M_2}{M_1}} \, 
\epsilon_\mu(v, \mp 1) \,, \nonumber\\
& &\bar u_2\Big(v, \pm \frac{1}{2}\Big)
\gamma_5 \not\!\bar\epsilon^{\,\ast}(\mp 1) \, 
u_\mu\Big(v, \pm \frac{3}{2}\Big) = 0 \,, \nonumber\\
& &\bar u_2\Big(v, \pm \frac{1}{2}\Big) 
\gamma_5 \not\!\bar\epsilon^{\,\ast}(\pm 1)\, 
u_\mu\Big(v, \mp \frac{1}{2}\Big) = \pm \frac{4}{\sqrt{3}} 
\ \sqrt{M_1 M_2} \, 
\epsilon_\mu(v, 0) \,. 
\en  
In the calculation of the helicity amplitudes 
$H_{\pm \frac{1}{2} \pm 1}$ and $H_{\pm \frac{1}{2} \mp 1}$ one can make use 
of the HQL identities  
\eq 
& &\bar u_2\Big(v, \pm \frac{1}{2}\Big)
\gamma_5 \not\! q \, \bar\epsilon^{\,\ast\mu}(\mp 1) 
u_\mu\Big(v, \pm \frac{3}{2}\Big) 
= \sqrt{3} \, \bar u_2\Big(v, \pm \frac{1}{2}\Big) 
\gamma_5 \not\! q \, \bar\epsilon^{\,\ast\mu}(\pm 1) 
u_\mu\Big(v, \mp \frac{1}{2}\Big) 
= \pm M_+ M_- \ \sqrt{\frac{M_2}{M_1}} \,, \nonumber\\ 
& &\bar u_2\Big(v, \pm \frac{1}{2}\Big)
\gamma_5 \not\!\bar\epsilon^{\,\ast}(\pm 1) \, q^\mu 
u_\mu\Big(v, \mp \frac{1}{2}\Big) = \pm \frac{2}{\sqrt{3}} \, 
 M_+ M_- \ \sqrt{\frac{M_2}{M_1}}
\en  

\section{Nonrelativistic quark model: spin--flavor wave 
functions, radiative decay constants and widths of double heavy baryons} 

In this Appendix we present results on the radiative decay amplitudes 
and widths of the DHBs in the nonrelativistic quark model. 
As emphasized before 
the nonrelativistic quark model is based on the spin--flavor wave functions 
which arise in the nonrelativistic limit of the relativistically covariant 
double--heavy three--quark currents with quantum numbers $J^{P}=\frac{1}{2}^+$ 
and $\frac{3}{2}^+$. The corresponding quark model 
spin--flavor wave functions are given in Table 2, where we use the following
notation for the antisymmetric 
$\chi_{_{A}}(\lambda)$ and symmetric $\chi_{_{S}}(\lambda)$, 
$\chi_{_{S}}^\ast(\lambda)$  
spin wave functions where $\lambda$ is the helicity of the baryon state: 
\eq\label{spin_wf}  
\chi_{_{A}}\Big(\frac{1}{2}\Big) &=& \sqrt{\frac{1}{2}} \ \biggl\{ 
\uparrow (\uparrow \downarrow - \downarrow \uparrow) \biggr\}\,, 
\quad\quad  \hspace*{1.5cm}
\chi_{_{A}}\Big(-\frac{1}{2}\Big) \ = \ \sqrt{\frac{1}{2}} \ \biggl\{ 
\downarrow (\uparrow \downarrow - \downarrow \uparrow) \biggr\}\,, 
\nonumber\\
\chi_{_{S}}\Big(\frac{1}{2}\Big) &=&  \sqrt{\frac{1}{6}}  \ \biggl\{ 
 2 \downarrow \uparrow \uparrow 
-  \uparrow (\uparrow \downarrow 
+  \downarrow \uparrow)  \biggr\} 
\,, \quad\quad \hspace*{.3cm} 
\chi_{_{S}}\Big(-\frac{1}{2}\Big) \ = \ - \sqrt{\frac{1}{6}}  \ \biggl\{ 
2 \uparrow \downarrow \downarrow 
- \downarrow (\uparrow \downarrow 
+ \downarrow \uparrow) \biggr\} \,, \\  
\chi_{_{S}}^\ast\Big(\frac{3}{2}\Big) &=& \uparrow \uparrow \uparrow\,, 
\quad\quad \hspace*{3.95cm} 
\chi_{_{S}}^\ast\Big(-\frac{3}{2}\Big) \ = \ \downarrow \downarrow \downarrow 
\,, \nonumber\\
\chi_{_{S}}^\ast\Big(\frac{1}{2}\Big) &=& \sqrt{\frac{1}{3}}  \ \biggl\{ 
  \uparrow \uparrow \downarrow + \uparrow \downarrow \uparrow   
+ \downarrow \uparrow \uparrow \biggr\} \,, \,, \quad\quad  \hspace*{.5cm}
\chi_{_{S}}^\ast\Big(-\frac{1}{2}\Big) \ = \ \sqrt{\frac{1}{3}}  \ \biggl\{ 
  \uparrow \downarrow \downarrow + \downarrow \uparrow \downarrow   
+ \downarrow \downarrow \uparrow \biggr\} \,.  \nonumber
\en 
In (\ref{spin_wf}) we use the ordering $\{q\,Q_{1}Q_{2}\}$.  
Next we relate the DHB radiative decay amplitudes to the nonrelativistic
amplitudes $G_{\lambda_2\lambda_\gamma}$, where  
$\lambda_\gamma = \pm 1$ and $\lambda_2 = \pm 1/2, \pm 3/2$ are 
the helicity components of the on--shell photon and the daughter baryon, 
respectively (see details in Appendix A). In order to evaluate the spin flip 
matrix elements between the baryon states we make use of the spin--flip 
(spin raising/lowering) operator 
\eq 
{\cal S_{\rm flip}^{\pm}} = 
- \sqrt{2} \, \sum\limits_{i=1}^3 \, \mu_i (\sigma_\pm)_i \,, 
\en 
where $i$ runs over the three constituent quarks.
The spin flip matrix elements are given by
\eq 
G_{\lambda_2 \mp 1} = 
\la B_2(\lambda_2)| {\cal S_{\rm flip}^\pm} |B_1(\lambda_1) \ra \,, 
\en 
where $\lambda_1 = \lambda_2 - \lambda_\gamma$ is the helicity of 
the parent baryon and $\sigma_\pm = \sigma_1 \pm i \sigma_2$.  
$\mu_i = e_i/(2m_i)$ is the $i$-th quark magnetic moment and 
$e_i$ and $m_i$ are its charge and mass, respectively. 

For the amplitude $G_{\lambda_2 \lambda_\gamma}$ one obtains 

Transition $\frac{1}{2}^A \to \frac{1}{2}^S$\,:
\eq 
G_{\pm \frac{1}{2}\pm 1} = - \sqrt{\frac{2}{3}} \, (\mu_c - \mu_b) \,. 
\en 
Transition $\frac{3}{2} \to \frac{1}{2}^A$\,:
\eq 
G_{\pm \frac{1}{2}\mp 1} =  - \sqrt{3} \, G_{\pm \frac{1}{2}\pm 1} 
= \pm (\mu_c - \mu_b) \,. 
\en
Transition $\frac{3}{2} \to \frac{1}{2}^S$\,:
\eq 
G_{\pm \frac{1}{2}\mp 1} =  - \sqrt{3} \, G_{\pm \frac{1}{2}\pm 1} =
\pm \frac{2}{\sqrt{3}} 
\ \biggl( \mu_q - \frac{\mu_{Q_1} + \mu_{Q_2}}{2} \biggr)\,. 
\en
Transition $\frac{1}{2}^{h} \to \frac{1}{2}^{l}$\,:
\eq 
G_{\pm \frac{1}{2}\pm 1} = \frac{2\sqrt{2}}{3} \, \sin\!2\theta_B    
\, \biggl( \mu_q 
- \frac{\sqrt{3}}{2} \, (\mu_c - \mu_b) \, \cot\! 2\theta_B \biggr) \,.
\en 
Transition $\frac{3}{2} \to \frac{1}{2}^{l}$\,:
\eq 
G_{\pm \frac{1}{2}\mp 1} =  - \sqrt{3} \, G_{\pm \frac{1}{2}\pm 1} 
= \pm \frac{2}{\sqrt{3}} \, \cos\theta_B \, 
\biggl( \mu_q - \frac{\mu_c}{2} \tan_+
              - \frac{\mu_b}{2} \tan_- \biggr) \,. 
\en
Transition $\frac{3}{2} \to \frac{1}{2}^{h}$\,:
\eq 
G_{\pm \frac{1}{2}\mp 1} =  - \sqrt{3} \, G_{\pm \frac{1}{2}\pm 1} 
= \pm \frac{2}{\sqrt{3}} \, \sin\theta_B \, 
\biggl( \mu_q - \frac{\mu_c}{2} \cot_- 
              - \frac{\mu_b}{2} \cot_+ \biggr) \,, 
\en
where $\tan_\pm = 1 \pm \tan\theta_B \sqrt{3}$ and 
      $\cot_\pm = 1 \pm \cot\theta_B \sqrt{3}$.

The states $\frac{1}{2}^A$, $\frac{1}{2}^S$, $\frac{1}{2}^{l}$ 
and $\frac{1}{2}^{h}$ 
correspond to baryons with antisymmetric, symmetric and 
mixed heavy quark spin configuration, respectively. In particular, 
$\frac{1}{2}^{h}$ and $\frac{1}{2}^{l}$ correspond to the mixed states $B_h$ 
and $B_l$.  
The index $q$ corresponds to the light $u, d$ or $s$ quark. 

In terms of the nonrelativistic amplitude $G_{\lambda_2\lambda_\gamma}$ the 
helicity 
amplitudes $H_{\lambda_2\lambda_\gamma}$ (see details in Appendix A) 
for the $\frac{1}{2}^+ \to \frac{1}{2}^+$ 
and $\frac{3}{2}^+ \to \frac{1}{2}^+$ radiative transitions are defined by: 
\eq 
H_{\lambda_2\lambda_\gamma} = |{\bf p}_2| \, \sqrt{N_1 N_2} \, 
G_{\lambda_2\lambda_\gamma} \,,  
\en 
where $N_i = \sqrt{2 M_i}$ is the extra factor acquired in the nonrelativistic 
quark model due to different normalization of states in the 
relativistic and the nonrelativistic theory. 
Therefore, we have 
\eq 
H_{\lambda_2\lambda_\gamma} = M_+ M_- \ \sqrt{\frac{M_2}{M_1}} \  
G_{\lambda_2\lambda_\gamma} \,.  
\en 
The radiative decay widths for the four possible $s_1 \to s_2$ spin 
transitions are given by:  
\eq\label{rates_NQM}  
& &\Gamma_{\frac{1}{2}^A \to \frac{1}{2}^S} = 
   \Gamma_{\frac{3}{2} \to \frac{1}{2}^A}   = 
   K \, \mu_c^2 \, \biggl(1 - \frac{\mu_b}{\mu_c} \biggr)^2  \,, 
\nonumber\\[2mm]
& &\Gamma_{\frac{1}{2}^{h} \to \frac{1}{2}^{l}} = \frac{4}{3} 
\, K \, \mu_q^2 \sin^2\!2\theta_B \biggl( 1 - 
\frac{\sqrt{3}}{2} \, \frac{\mu_c - \mu_b}{\mu_q} \, \cot 2\theta_B  
\biggr)^2 \,, \nonumber\\
& &\Gamma_{\frac{3}{2} \to \frac{1}{2}^S} = \frac{4}{3} \, K \, 
\mu_q^2 \, \biggl(1 - \frac{\mu_{Q_1}  + \mu_{Q_1}}{2 \mu_q} \biggr)^2  
\,, \nonumber\\[2mm] 
& &\Gamma_{\frac{3}{2} \to \frac{1}{2}^{l}} = \frac{4}{3} \, K \, 
\mu_q^2 \, \cos^2\!\theta_B \, \biggl(1 - \frac{\mu_c}{2\mu_q} \tan_+ 
- \frac{\mu_b}{2\mu_q} \tan_-
\biggr)^2  \,, \\[2mm] 
& &\Gamma_{\frac{3}{2} \to \frac{1}{2}^{h}} = \frac{4}{3} \, K \, 
\mu_q^2 \, \sin^2\!\theta_B \, 
\biggl(1 - \frac{\mu_c}{2\mu_q} \cot_- 
- \frac{\mu_b}{2\mu_q} \cot_+ \biggr)^2  \,, \nonumber 
\en
where 
\eq 
K = \alpha \, M_2 \, \frac{(M_1^2-M_2^2)^3}{6 M_1^4}  
\en 
and $\alpha \simeq  1/137$ is the fine-structure constant. 

It is evident that the widths of the subleading processes  
$\frac{1}{2}^A \to \frac{1}{2}^S$ and 
$\frac{3}{2} \to \frac{1}{2}^A$ are suppressed by a factor of $(m_q/m_c)^2$
compared to the widths of the leading process $\frac{3}{2} \to \frac{1}{2}^S$.   

\section{Mass operator and radiative vertex functions of double heavy baryons}

The baryon mass operators $\Sigma_{B}(p)$  
and $\Sigma_{B^\ast}^{\mu\nu}(p)$ are given by
\eq\label{mass_operator_B}
\Sigma_{B}(p) &=& 6 \, N_f \, \int dk_{123} \,\, \Phi^{2}(z_0)
\, R_{\Sigma}(k_1^+,k_2^+,k_3^+) \,, \nonumber\\ 
R_{\Sigma}(r_1,r_2,r_3) &=& \Gamma_{1f} \tilde S_{q}(r_3)
\overline\Gamma_{1i} \,
{\rm tr} \left[\Gamma_{2f} \tilde S_{Q_2}(r_2) 
\overline\Gamma_{2i} \tilde S_{Q_1}(-r_1)\right]  
\en 
\eq\label{mass_operator_Bstar}
\Sigma_{B^\ast}^{\mu\nu}(p) &=&  6 \, \int dk_{123} \,\, \Phi^{2}(z_0)
\, R_{\Sigma}^{\mu\nu}(k_1^+,k_2^+,k_3^+) \,, \nonumber\\ 
R_{\Sigma}^{\mu\nu}(r_1,r_2,r_3) &=& \Gamma_{1f} 
\tilde S_{q}(r_3)\overline\Gamma_{1i} \,
{\rm tr} \left[\Gamma_{2f}^\mu \tilde S_{Q_2}(r_2)
\overline\Gamma_{2i}^\nu \tilde S_{Q_1}(-r_1)\right] \,.  
\en 
Here and in the following  
$N_f$ denotes a statistical flavor factor, which is equal to 1 or 2 for DHBs  
with two different or two identical heavy quarks, respectively, and 
$\overline\Gamma = \gamma^0 \Gamma^\dagger \gamma^0$. We have introduced the
abbreviations
\eq 
dk_{123} &=& \frac{d^{4}k_{1}d^{4}k_{2}d^{4}k_{3}}{(2\pi)^8 i^2}
\, \delta^{4}(k_{1}+k_{2}+k_{3})  \,, \quad  
z_0 \, = \,  - \frac{1}{3} \, (k_{1}^{2}+k_{2}^{2}+k_{3}^{2})\,, \nonumber\\
k_i^+ &=& k_i + p w_i \,, \quad  
k_i^{\prime \, +} \, = \, k_i + p^\prime w_i\,, \quad 
L_i \, = \, \frac{2}{3} \Big(k_i - \sum\limits_{j=1}^3 k_j w_j\Big) \,, 
\nonumber\\[1mm]
z_1(q) &=& L_1 q - \frac{2}{3} q^{2}(w_{2}^{2}+w_{2}w_{3}+w_{3}^{2})\,, 
\nonumber\\[2mm]
z_2(q) &=& L_2 q - \frac{2}{3} q^{2}(w_{1}^{2}+w_{1}w_{3}+w_{3}^{2})\,, 
\nonumber\\[2mm]
z_3(q) &=& L_3 q - \frac{2}{3} q^{2}(w_{1}^{2}+w_{1}w_{2}
+w_{2}^{2}) 
\en 
and 
\eq
R_{\mu\,, \, \Delta_1}^\perp(r_1,r_2,r_3,q) &=&
- \Gamma_{1f} \tilde S_{q}(r_3)\overline\Gamma_{1i} \,
{\rm tr} \left[\Gamma_{2f} \tilde S_{Q_2}(r_2)\overline\Gamma_{2i}
\tilde S_{Q_1}(-r_1)\gamma^{\perp}_{\mu} 
\tilde S_{Q_1}(-r_1 + q)\right] \,, \nonumber\\[3mm]
R_{\mu\,, \, \Delta_2}^\perp(r_1,r_2,r_3,q) &=&
\Gamma_{1f} \tilde S_{q}(r_3)\overline\Gamma_{1i} \,
{\rm tr} \left[\Gamma_{2f} \tilde S_{Q_2}(r_2 - q)
\gamma^{\perp}_{\mu} \tilde S_{Q_2}(r_2)\overline\Gamma_{2i}
\tilde S_{Q_1}(-r_1)\right] \,, \nonumber\\[3mm]
R_{\mu \,, \, \Delta_3}^\perp(r_1,r_2,r_3,q) &=&
\Gamma_{1f} \tilde S_{q}(r_3 - q)\gamma^{\perp}_{\mu} 
\tilde S_{q}(r_3)\overline\Gamma_{1i}
{\rm tr} \left[\Gamma_{2f} \tilde S_{Q_2}(r_2)
\overline\Gamma_{2i} \tilde S_{Q_1}(-r_1)\right] \,, \nonumber\\[3mm]
R_{\mu\nu, \, \Delta_1}^\perp(r_1,r_2,r_3,q) &=&
- \Gamma_{1f} \tilde S_{q}(r_3)\overline\Gamma_{1i} \,
{\rm tr} \left[\Gamma_{2f} \tilde S_{Q_2}(r_2)\overline\Gamma_{2i, \nu}
\tilde S_{Q_1}(-r_1)\gamma^{\perp}_{\mu} 
\tilde S_{Q_1}(-r_1 + q)\right] \,, \nonumber\\[5mm]
R_{\mu\nu, \, \Delta_2}^\perp(r_1,r_2,r_3,q) &=&
\Gamma_{1f} \tilde S_{q}(r_3)\overline\Gamma_{1i} \,
{\rm tr} \left[\Gamma_{2f} \tilde S_{Q_2}(r_2 - q)
\gamma^{\perp}_{\mu} \tilde S_{Q_2}(r_2)\overline\Gamma_{2i, \nu}
\tilde S_{Q_1}(-r_1)\right] \,, \nonumber\\[5mm]
R_{\mu\nu, \, \Delta_3}^\perp(r_1,r_2,r_3,q) &=&
\Gamma_{1f} \tilde S_{q}(r_3 - q)\gamma^{\perp}_{\mu} 
\tilde S_{q}(r_3)\overline\Gamma_{1i}
{\rm tr} \left[\Gamma_{2f} \tilde S_{Q_2}(r_2)
\overline\Gamma_{2i, \nu} \tilde S_{Q_1}(-r_1)\right] \,, \nonumber\\[5mm]
R_{\nu, \Sigma}(r_1,r_2,r_3) &=& 
\Gamma_{1f} \tilde S_{q}(r_3)\overline\Gamma_{1i} \,
{\rm tr} \left[\Gamma_{2f} \tilde S_{Q_2}(r_2)\overline\Gamma_{2i, \nu} 
\tilde S_{Q_1}(-r_1)\right] \,. 
\en 
In the following we present explicit expressions for 
the electromagnetic vertex function. 
In case of the $(\frac{1}{2}^+, 0) \to (\frac{1}{2}^+, 1)$ and 
$(\frac{3}{2}^+, 1) \to (\frac{1}{2}^+, 0)$ 
transitions the expressions for the nonvanishing contribution  
of the triangle diagrams in Figs.2(a)-2(c) 
(terms $\Lambda_{\mu, \, \Delta}^{\perp}$ and 
$\Lambda_{\mu\nu, \, \Delta}^{\perp}$, respectively) read 
\seq\label{elements11}
\eq
\Lambda_{\mu, \, \Delta}^{\perp}(p_{1},p_{2})
&=& 6 \, N_f \, g_{B_1} \, g_{B_2} \, \int dk_{123} \,\, \sum\limits_{i=1}^3 \,
e_{i} \, \Phi(z_0) \, \Phi\Big(z_0 + z_i(q)\Big)
\, R_{\mu \,, \Delta_i}^\perp(k_1^+,k_2^+,k_3^+,q)\,,\\ 
\Lambda_{\mu\nu, \, \Delta}^{\perp}(p_{1},p_{2})
&=& 6 \, N_f \, 
g_{B_1^\ast} \, g_{B_2} \, \int dk_{123} \,\, \sum\limits_{i=1}^3 \,
e_{i} \, \Phi(z_0) \, \Phi\Big(z_0 + z_i(q)\Big)
\, R_{\mu\nu, \, \Delta_i}^\perp(k_1^+,k_2^+,k_3^+,q)\,,
\en
\sen 
where $e_1=e_{Q_1}$, $e_2=e_{Q_2}$ and $e_3=e_q$.  
For the $(\frac{3}{2}^+, 1) \to (\frac{1}{2}^+, 1)$ 
transition the electromagnetic 
vertex function $\Lambda_{\mu\nu}$ obtains contributions 
from the triangle diagram $\Lambda_{\mu\nu, \, \Delta}$ [Fig.2(a)-2(c)], 
the left and right bubble diagrams $\Lambda_{\mu\nu, {\rm bub}_L}^{\perp}$ 
[Fig.2(d)] and 
$\Lambda_{\mu\nu, {\rm bub}_R}^{\perp}$ [Fig.2(e)], and the pole diagram 
$\Lambda_{\mu\nu, {\rm pol}_L}^{\perp}$ [Fig.2(f)]. 
The corresponding contributions are given by 
\seq\label{elements31}
\eq
\Lambda_{\mu\nu, \, {\rm bub}_L}^{\perp}(p_{1},p_{2})
&=& - 6 \, N_f \, g_{B_1^\ast} \, g_{B_2} \, 
\int dk_{123} \,\, \sum\limits_{i=1}^3 \,
e_{i} \, L_{i \mu}^\perp \, \Phi(z_0) \, \int\limits_0^1 \, dt \,
\Phi^\prime\Big(z_0 + t z_i(-q)\Big)
\, R_{\nu, \Sigma}(k_1^{\prime \, +},k_2^{\prime \, +},k_3^{\prime \, +}) \,,\\
\Lambda_{\mu\nu, \, {\rm bub}_R}^{\perp}(p_{1},p_{2})
&=& - 6 \, N_f \, g_{B_1^\ast} \, g_{B_2} \, 
\int dk_{123} \,\, \sum\limits_{i=1}^3 \,
e_{i} \, L_{i \mu}^\perp \, \Phi(z_0) \, \int\limits_0^1 \, dt \,
\Phi^\prime\Big(z_0 + t z_i(q)\Big) \, R_{\nu, \Sigma}(k_1^+,k_2^+,k_3^+) \,,\\
\Lambda_{\mu\nu, \, {\rm pol}_L}^{\perp}(p_{1},p_{2})
&=& 6 \, N_f \, g_{B_1^\ast} \, g_{B_2} \, \int dk_{123} \, \, \Phi^2(z_0)
\ R_{\alpha, \Sigma}(k_1^{\prime \, +},k_2^{\prime \, +},k_3^{\prime \, +}) \ 
\tilde S_{B^\ast}^{\alpha\beta}(p_{2}) \, 
\Big( g_{\mu\nu}^{\perp} \gamma_\beta - \gamma_\mu^\perp  g_{\nu\beta} \Big)
\,. 
\en
\sen

\newpage 

\newpage 

\begin{figure}
\centering{\
\epsfig{figure=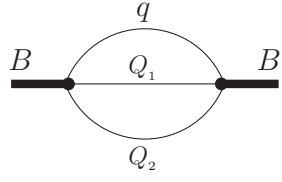,scale=.8}}
\caption{Diagram describing the double heavy baryon mass operator.}
\label{fig:str}
\end{figure}

\vspace*{2cm} 

\begin{figure}
\centering{\
\epsfig{figure=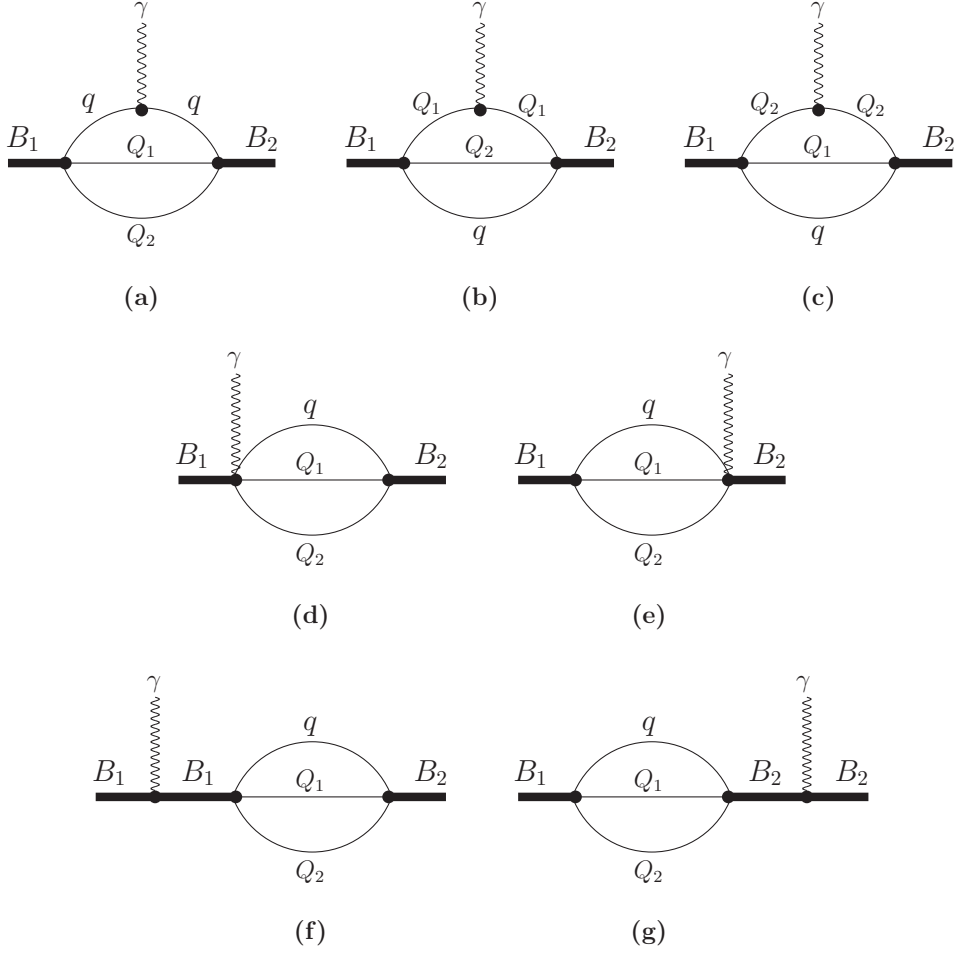,scale=.8}}
\caption{Diagrams contributing to the
radiative transitions of double heavy baryons.} 
\label{fig:rad_trans}
\end{figure}

\newpage 

\begin{table}
\begin{center}
{\bf Table I.} 
Classification and mass values of double--heavy baryons. Mass values are based
on~\cite{Ebert:2004ck} \\
except for the $\Xi_{cc}$ mass which is taken from~\cite{Amsler:2008zzb}.  

\vspace*{.25cm}
\def\arraystretch{1.1}
\begin{tabular}{|c|c|c|c|c|}
\hline 
\,\, Notation \,\,  & \,\, Content \,\, &  \,\, $J^P$ \,\, &
\,\, $S_d$ \,\, & \,\, Mass (GeV)\,\, \\[2mm]
\hline
$\Xi_{cc}$  & $q\{cc\}$ & $1/2^+$ & $1$ & $3.5189$ \\
\hline
$\Xi_{bc}$  & $q\{bc\}$ & $1/2^+$ & $1$ & $6.933$ \\ 
\hline
$\Xi'_{bc}$ & $q[bc]$   & $1/2^+$ & $0$ & $6.963$ \\
\hline
$\Xi_{bb}$  & $q\{bb\}$ & $1/2^+$ & $1$ & $10.202$ \\
\hline
$\Xi_{cc}^\ast$  & $q\{cc\}$ & $3/2^+$ & $1$ & $3.727$ \\
\hline
$\Xi_{bc}^\ast$  & $q\{bc\}$ & $3/2^+$ & $1$ & $6.980$ \\
\hline
$\Xi_{bb}^\ast$  & $q\{bb\}$ & $3/2^+$ & $1$ & $10.237$ \\
\hline
$\Omega_{cc}$  & $s\{cc\}$ & $1/2^+$ & $1$ & $3.778$ \\
\hline
$\Omega_{bc}$  & $s\{bc\}$ & $1/2^+$ & $1$ & $7.088$ \\ 
\hline
$\Omega'_{bc}$ & $s[bc]$   & $1/2^+$ & $0$ & $7.116$ \\
\hline
$\Omega_{bb}$  & $s\{bb\}$ & $1/2^+$ & $1$ & $10.359$ \\
\hline
$\Omega_{cc}^\ast$  & $s\{cc\}$ & $3/2^+$ & $1$ & $3.872$ \\
\hline
$\Omega_{bc}^\ast$  & $s\{bc\}$ & $3/2^+$ & $1$ & $7.130$ \\
\hline
$\Omega_{bb}^\ast$  & $s\{bb\}$ & $3/2^+$ & $1$ & $10.389$ \\
\hline
\end{tabular}
\end{center}
\end{table}

\vspace*{1cm}

\begin{table}
\begin{center} 
{\bf Table II.} 
DHB wave functions. 

\vspace*{.25cm}
\def\arraystretch{1.1}
\begin{tabular}{|c|c||c|c|} 
\hline
  $\;\;$ Baryon $\;\;$ & $\;\;\;\;\;\;\;$ Wave function $\;\;\;\;\;\;\;$ &
  $\;\;$ Baryon $\;\;$ & $\;\;\;\;\;\;\;$ Wave function $\;\;\;\;\;\;\;$ \\ 
\hline 
$\Xi_{cc}$ & $qcc \,\,\, \chi_{_{S}}(\lambda)$ &
$\Omega_{cc}$ & $scc \,\,\, \chi_{_{S}}(\lambda)$ \\
\hline 
$\Xi_{bb}$ & $qbb \,\,\, \chi_{_{S}}(\lambda)$ &
$\Omega_{bb}$ & $sbb \,\,\, \chi_{_{S}}(\lambda)$ \\
\hline 
$\Xi_{bc}$ & $\frac{1}{\sqrt{2}} q (bc + cb)  \,\,\, \chi_{_{S}}(\lambda)$ & 
$\Omega_{bc}$ & $\frac{1}{\sqrt{2}} s (bc + cb) \,\,\, \chi_{_{S}}(\lambda)$ \\
\hline 
$\Xi'_{bc}$ & $\frac{1}{\sqrt{2}} q (bc - cb)  \,\,\, \chi_{_{A}}(\lambda)$ &
$\Omega'_{bc}$ & $\frac{1}{\sqrt{2}} s (bc - cb) \,\,\, \chi_{_{A}}(\lambda)$ 
\\
\hline 
$\Xi^\ast_{cc}$ & $- qcc \,\,\, \chi_{_{S}}^\ast(\lambda)$ & 
$\Omega^\ast_{cc}$ & $- scc \,\,\, \chi_{_{S}}^\ast(\lambda)$ \\
\hline 
$\Xi^\ast_{bb}$ & $- qbb \,\,\, \chi_{_{S}}^\ast(\lambda)$ & 
$\Omega^\ast_{bb}$ & $- sbb \,\,\, \chi_{_{S}}^\ast(\lambda)$ \\
\hline 
$\Xi^\ast_{bc}$ & $- \frac{1}{\sqrt{2}} q (bc+cb) \,\,\, 
\chi_{_{S}}^\ast(\lambda)$ & 
$\Omega^\ast_{bc}$ & $- \frac{1}{\sqrt{2}} s (bc+cb) \,\,\, 
\chi_{_{S}}^\ast(\lambda)$ \\
\hline
\end{tabular}
\end{center} 
\end{table}

\newpage 

\begin{table}
\begin{center}
{\bf Table III.} 
Radiative decay widths of DHBs in keV. 

\vspace*{.25cm}
\def\arraystretch{1.2}
\begin{tabular}{|c|c|c|c|c|}
\hline 
\,\, Decay mode \,\, & \,\, Exact results \,\, & 
\,\, HQL \,\, &  
\,\, NQM \,\, & 
NQM + HQL \\[2mm]
\hline
$\Xi_{bc}^{' +}\to \Xi_{bc}^+$        
& (1.56 $\pm$ 0.08) $\times$ 10$^{-2}$ 
& 0 
& 1.35 $\times$ 10$^{-2}$ & 0 \\
\hline
$\Xi_{bc}^{' 0}\to \Xi_{bc}^0$        
& (1.56 $\pm$ 0.08) $\times$ 10$^{-2}$ 
& 0
& 1.35 $\times$ 10$^{-2}$ & 0 \\
\hline 
$\Omega_{bc}' \to \Omega_{bc}$ 
& (1.26 $\pm$ 0.05) $\times$ 10$^{-2}$
& 0
& 1.10 $\times$ 10$^{-2}$ & 0 \\
\hline 
$\Xi_{bc}^{h +}\to \Xi_{bc}^{l +}$        
& 0.14 $\pm$ 0.03  
& $\simeq$ 0.17 
& 0.34 & 0.53 \\
\hline 
$\Xi_{bc}^{h 0}\to \Xi_{bc}^{l 0}$        
& 0.31 $\pm$ 0.04
& $\simeq$ 0.04
& 0.26 & 0.13 \\
\hline 
$\Omega_{bc}^{h} \to \Omega_{bc}^{l}$ 
& 0.21 $\pm$ 0.02 
& $\simeq$ 0.02 
& 0.15 
& 0.06 \\
\hline 
$\Xi_{bc}^{\ast^+}\to \Xi_{bc}^{' +}$        
& (0.28 $\pm$ 0.01) $\times$ 10$^{-2}$
& 0 
& 0.25 $\times$ 10$^{-2}$ & 0 \\
\hline 
$\Xi_{bc}^{\ast^0}\to \Xi_{bc}^{' 0}$        
& (0.28 $\pm$ 0.01) $\times$ 10$^{-2}$
& 0 
& 0.25 $\times$ 10$^{-2}$ & 0 \\
\hline 
$\Omega_{bc}^\ast \to \Omega_{bc}'$        
& (0.16 $\pm$ 0.01) $\times$ 10$^{-2}$
& 0 
& 0.14 $\times$ 10$^{-2}$ & 0 \\ 
\hline
$\Xi^{\ast ++}_{cc} \to \Xi^{++}_{cc}$       
& 23.46 $\pm$ 3.33
& 20.53 $\pm$ 0.79 
& 36.22 & 63.88 \\
\hline 
$\Xi^{\ast +}_{cc} \to \Xi^{+}_{cc}$       
& 28.79 $\pm$ 2.51
& 5.13 $\pm$ 0.20  
& 35.65 & 15.97 \\
\hline 
$\Omega^\ast_{cc} \to \Omega_{cc}$ 
& 2.11 $\pm$ 0.11 
& $\simeq$ 0.29  
& 2.42 & 0.87 \\
\hline 
$\Xi^{\ast +}_{bc} \to \Xi^+_{bc}$       
& 0.49 $\pm$ 0.09
& $\simeq$ 0.27   
& 0.67 & 0.83 \\ 
\hline 
$\Xi^{\ast 0}_{bc} \to \Xi^0_{bc}$       
& 0.24 $\pm$ 0.04
& $\simeq$ 0.07   
& 0.30 & 0.21 \\ 
\hline 
$\Omega^\ast_{bc} \to \Omega_{bc}$ 
& 0.12 $\pm$ 0.02
& $\simeq$ 0.03 
& 0.13 & 0.08 \\ 
\hline 
$\Xi^{\ast +}_{bc} \to \Xi^{l +}_{bc}$       
& 0.46 $\pm$ 0.10
& $\simeq$ $0.37$ 
& 0.69 & 1.14 \\ 
\hline 
$\Xi^{\ast +}_{bc} \to \Xi^{h +}_{bc}$       
& (0.15 $\pm$ 0.02) $\times$ 10$^{-2}$
& $\simeq$ 0.03 $\times$ 10$^{-2}$ 
& 0.16 $\times$ 10$^{-2}$ 
& 0.08 $\times$ 10$^{-2}$ \\ 
\hline 
$\Xi^{\ast 0}_{bc} \to \Xi^{l 0}_{bc}$       
& 0.51 $\pm$ 0.06 
& $\simeq$ 0.10
& 0.59 & 0.28 \\ 
\hline 
$\Xi^{\ast 0}_{bc} \to \Xi^{h 0}_{bc}$       
& (0.02 $\pm$ 0.02) $\times$ 10$^{-4}$
& $\simeq$ 0.06 $\times$ 10$^{-3}$ 
& 0.01 $\times$ 10$^{-3}$ & 0.19 $\times$ 10$^{-3}$ \\
\hline 
$\Omega^\ast_{bc} \to \Omega_{bc}^{l}$ 
& (0.29 $\pm$ 0.03) 
& $\simeq$ 0.03 
& 0.30 
& 0.12 \\ 
\hline 
$\Omega^\ast_{bc} \to \Omega_{bc}^{h}$ 
& (0.01 $\pm$ 0.01) $\times$ 10$^{-4}$
& $\simeq$ 0.01 $\times$ 10$^{-3}$  
& 0.01 $\times$ 10$^{-4}$ 
& 0.03 $\times$ 10$^{-3}$ \\ 
\hline 
$\Xi^{\ast 0}_{bb} \to \Xi^0_{bb}$       
& 0.31 $\pm$ 0.06
& $\simeq$ 0.11 
& 0.38 & 0.35 \\ 
\hline 
$\Xi^{\ast -}_{bb} \to \Xi^-_{bb}$       
& (5.87 $\pm$ 1.42) $\times$ 10$^{-2}$
& $\simeq$ 2.8 $\times$ 10$^{-2}$ 
& 7.34 $\times$ 10$^{-2}$ 
& 8.69 $\times$ 10$^{-2}$ \\
\hline 
$\Omega^\ast_{bb} \to \Omega_{bb}$ 
& (2.26 $\pm$ 0.45) $\times$ 10$^{-2}$
& $\simeq$ 1.0 $\times$ 10$^{-2}$  
& 2.36 $\times$ 10$^{-2}$ 
& 2.97 $\times$ 10$^{-2}$ \\
\hline 
\end{tabular}
\end{center}
\end{table}  

\newpage

\begin{table}
\begin{center}
{\bf Table IV.} 
$\theta_B$ dependence of radiative decay widths involving mixed DHBs in eV.

\vspace*{.25cm}
\def\arraystretch{1.1}
\begin{tabular}{|c|c|c|c|c|c|}
\hline 
\,\, $\theta_B$ \,\, 
& \,\, Decay mode \,\, & \,\, Exact results \,\, & 
\,\, HQL \,\, &  \,\, NQM \,\, & NQM + HQL \\[2mm]
\hline 
\hline 
& $\Xi_{bc}^{h +}\to \Xi_{bc}^{l +}$        
& 0.2 $\pm$ 0.2 
& $\simeq$ 10 
& 3
& 31 \\ 
\cline{2-6}
& $\Xi_{bc}^{h 0}\to \Xi_{bc}^{l 0}$        
& 170 $\pm$ 15 
& $\simeq$ 3
& 43
& 8 \\
\cline{2-6}
& $\Omega_{bc}^{h} \to \Omega_{bc}^{l}$ 
& 130 $\pm$ 10
& $\simeq$ 1
& 28
& 3 \\
\cline{2-6}
\cline{2-6}
& $\Xi^{\ast +}_{bc} \to \Xi^{h +}_{bc}$       
& 0.07 $\pm$ 0.01 
& $\simeq$ 0.3 
& 5 
& 1 \\
\cline{2-6}
10$^0$ 
& $\Xi^{\ast 0}_{bc} \to \Xi^{h 0}_{bc}$       
& $\simeq$ 0.1  
& $\simeq$ 0.1
& 0.7 
& 0.3 \\
\cline{2-6} 
& $\Omega^\ast_{bc} \to \Omega_{bc}^{h}$ 
& $\simeq$ 0.03  
& $\simeq$ 0.02
& 0.5 
& 0.1 \\ 
\cline{2-6}
& $\Xi^{\ast +}_{bc} \to \Xi^{l +}_{bc}$       
& 687 $\pm$ 132
& 276 $\pm$ 11 
& 626
& 859 \\
\cline{2-6}
& $\Xi^{\ast 0}_{bc} \to \Xi^{l 0}_{bc}$       
& 475 $\pm$ 65 
& 69 $\pm$ 3 
& 360 
& 215 \\ 
\cline{2-6}
& $\Omega^\ast_{bc} \to \Omega_{bc}^{l}$ 
& 263 $\pm$ 26   
& $\simeq$ 28 
& 164
& 84 \\
\hline
\hline 
& $\Xi_{bc}^{h +}\to \Xi_{bc}^{l +}$        
& 16  $\pm$ 7  
& 27 $\pm$ 1
& 27 
& 84 \\ 
\cline{2-6}
& $\Xi_{bc}^{h 0}\to \Xi_{bc}^{l 0}$        
& 224 $\pm$ 24
& $\simeq$ 7 
& 73 
& 21 \\ 
\cline{2-6}
& $\Omega_{bc}^{h} \to \Omega_{bc}^{l}$ 
& 166 $\pm$ 14
& $\simeq$ 3
& 44
& 9 \\
\cline{2-6}
15$^0$ 
& $\Xi^{\ast +}_{bc} \to \Xi^{h +}_{bc}$       
& 0.9 $\pm$ 0.1    
& $\simeq$ 0.6 
& 6
& 2 \\ 
\cline{2-6}
& $\Xi^{\ast 0}_{bc} \to \Xi^{h 0}_{bc}$       
& $\simeq$ 0.05
& $\simeq$ 0.1 
& 0.2
& 0.5 \\
\cline{2-6}
& $\Omega^\ast_{bc} \to \Omega_{bc}^{h}$ 
& $\simeq$ 0.02
& $\simeq$ 0.1
& 0.2 
& 0.1 \\ 
\cline{2-6}
& $\Xi^{\ast +}_{bc} \to \Xi^{l +}_{bc}$       
& 619 $\pm$ 122   
& 289 $\pm$ 12 
& 619 
& 900 \\
\cline{2-6}
& $\Xi^{\ast 0}_{bc} \to \Xi^{l 0}_{bc}$       
& 493  $\pm$ 66   
& 72 $\pm$ 3
& 404 
& 225 \\ 
\cline{2-6}
& $\Omega^\ast_{bc} \to \Omega_{bc}^{l}$ 
& 276 $\pm$ 26   
& $\simeq$ 29 
& 187 
& 88 \\
\hline
\hline 
& $\Xi_{bc}^{h +}\to \Xi_{bc}^{l +}$        
& 63 $\pm$ 19   
& 64 $\pm$ 2 
& 100 
& 200 \\ 
\cline{2-6}
& $\Xi_{bc}^{h 0}\to \Xi_{bc}^{l 0}$        
& 272 $\pm$ 31   
& 16 $\pm$ 1 
& 126
& 51 \\
\cline{2-6}
& $\Omega_{bc}^{h} \to \Omega_{bc}^{l}$ 
& 195 $\pm$ 17   
& $\simeq$ 7.3
& 72 
& 22 \\
\cline{2-6}
20$^0$ 
& $\Xi^{\ast +}_{bc} \to \Xi^{h +}_{bc}$       
& 1.2 $\pm$ 0.1  
& $\simeq$ 1.5 
& 5
& 2 \\ 
\cline{2-6}
& $\Xi^{\ast 0}_{bc} \to \Xi^{h 0}_{bc}$       
& $\simeq$ 0.01   
& $\simeq$ 0.2
& 0.01 
& 0.5 \\
\cline{2-6}
& $\Omega^\ast_{bc} \to \Omega_{bc}^{h}$ 
& $\simeq$ 0.01 
& $\simeq$ 0.03 
& 0.03 
& 0.1 \\ 
\cline{2-6}
& $\Xi^{\ast +}_{bc} \to \Xi^{l +}_{bc}$       
& 546 $\pm$ 111   
& 313 $\pm$ 12 
& 632 
& 973 \\
\cline{2-6}
& $\Xi^{\ast 0}_{bc} \to \Xi^{l 0}_{bc}$       
& 504  $\pm$ 65   
& 78 $\pm$ 3 
& 468 
& 243 \\ 
\cline{2-6}
& $\Omega^\ast_{bc} \to \Omega_{bc}^{l}$ 
& 286 $\pm$ 27   
& $\simeq$ 32
& 223
& 96 \\ 
\hline
\hline 
& $\Xi_{bc}^{h +}\to \Xi_{bc}^{l +}$        
& 135 $\pm$ 34 
& 91 $\pm$ 4 
& 177 
& 284 \\ 
\cline{2-6}
& $\Xi_{bc}^{h 0}\to \Xi_{bc}^{l 0}$        
& 306 $\pm$ 37
& 223 $\pm$ 1 
& 142
& 71 \\ 
\cline{2-6}
& $\Omega_{bc}^{h} \to \Omega_{bc}^{l}$ 
& 213 $\pm$ 20   
& $\simeq$ 18 
& 131
& 52 \\
\cline{2-6}
25$^0$ 
& $\Xi^{\ast +}_{bc} \to \Xi^{h +}_{bc}$       
& 1.4 $\pm$ 0.2   
& $\simeq$ 0.3
& 2
& 1 \\ 
\cline{2-6}
& $\Xi^{\ast 0}_{bc} \to \Xi^{h 0}_{bc}$       
& 0.001 $\pm$ 0.001     
& $\simeq$ 0.08
& 0.004  
& 0.2 \\
\cline{2-6}
& $\Omega^\ast_{bc} \to \Omega_{bc}^{h}$ 
& 0.001 $\pm$ 0.001    
& $\simeq$ 0.02
& 0.002
& 0.04 \\ 
\cline{2-6}
& $\Xi^{\ast +}_{bc} \to \Xi^{l +}_{bc}$       
& 469 $\pm$ 99   
& 373 $\pm$ 15 
& 680 
& 1115 \\
\cline{2-6}
& $\Xi^{\ast 0}_{bc} \to \Xi^{l 0}_{bc}$       
& 507  $\pm$ 64   
& 90 $\pm$ 3 
& 578
& 278 \\ 
\cline{2-6}
& $\Omega^\ast_{bc} \to \Omega_{bc}^{l}$ 
& 291 $\pm$ 26    
& $\simeq$ 37 
& 281
& 111 \\
\hline 
\end{tabular}
\end{center}
\end{table}  

\begin{table}
\begin{center}
{\bf Table V.} 
Radiative decay widths of DHBs in keV. Comparison 
with the quark model~\cite{Albertus:2010hi}. 

\vspace*{.25cm}
\def\arraystretch{1.2}
\begin{tabular}{|c|c|c|}
\hline 
\,\, Decay mode \,\, & \,\, Quark model~\cite{Albertus:2010hi} \,\, & 
\, \, Our results \,\, \\[2mm]
\hline
$\Xi_{bc}^{' +}\to \Xi_{bc}^+$        
& 0.992 $\times$ 10$^{-2}$ 
& (1.56 $\pm$ 0.08) $\times$ 10$^{-2}$ \\
\hline
$\Xi_{bc}^{' 0}\to \Xi_{bc}^0$        
& 0.992 $\times$ 10$^{-2}$ 
& (1.56 $\pm$ 0.08) $\times$ 10$^{-2}$ \\
\hline 
$\Omega_{bc}' \to \Omega_{bc}$ 
& 3.69 $\times$ 10$^{-2}$ 
& (1.26 $\pm$ 0.05) $\times$ 10$^{-2}$ \\
\hline 
$\Xi_{bc}^{h +}\to \Xi_{bc}^{l +}$  
& 12.4 $\times$ 10$^{-2}$       
& $(14 \pm 4)$ $\times$ 10$^{-2}$ \\ 
\hline 
$\Xi_{bc}^{h 0}\to \Xi_{bc}^{l 0}$       
& 20.9 $\times$ 10$^{-2}$        
& (31 $\pm$ 4) $\times$ 10$^{-2}$ \\
\hline 
$\Omega_{bc}^{h} \to \Omega_{bc}^{l}$ 
& 8.52 $\times$ 10$^{-2}$        
& (21$\pm$ 2) $\times$ 10$^{-2}$ \\
\hline 
$\Xi_{bc}^{\ast^+}\to \Xi_{bc}^{' +}$ 
& 4.04 $\times$ 10$^{-2}$               
& (0.28 $\pm$ 0.01) $\times$ 10$^{-2}$ \\
\hline 
$\Xi_{bc}^{\ast^0}\to \Xi_{bc}^{' 0}$        
& 4.04 $\times$ 10$^{-2}$               
& (0.28 $\pm$ 0.01) $\times$ 10$^{-2}$ \\
\hline 
$\Omega_{bc}^\ast \to \Omega_{bc}'$      
& 3.69 $\times$ 10$^{-2}$                 
& (0.16 $\pm$ 0.01) $\times$ 10$^{-2}$ \\
\hline
$\Xi^{\ast +}_{bc} \to \Xi^+_{bc}$       
& 1.05
& 0.49 $\pm$ 0.09 \\
\hline 
$\Xi^{\ast 0}_{bc} \to \Xi^0_{bc}$       
& 0.505 
& 0.24 $\pm$ 0.04 \\
\hline 
$\Omega^\ast_{bc} \to \Omega_{bc}$ 
& 0.209
& 0.12 $\pm$ 0.02 \\
\hline 
$\Xi^{\ast +}_{bc} \to \Xi^{l +}_{bc}$       
& 0.739 
& 0.46 $\pm$ 0.10 \\
\hline 
$\Xi^{\ast +}_{bc} \to \Xi^{h +}_{bc}$   
& 6.05 $\times$ 10$^{-2}$     
& (0.15 $\pm$ 0.02) $\times$ 10$^{-2}$ \\
\hline 
$\Xi^{\ast 0}_{bc} \to \Xi^{l 0}_{bc}$       
& 1.03 
& 0.51 $\pm$ 0.06 \\
\hline 
$\Xi^{\ast 0}_{bc} \to \Xi^{h 0}_{bc}$   
& 0.12 $\times$ 10$^{-2}$     
& (0.02 $\pm$ 0.02) $\times$ 10$^{-4}$ \\
\hline 
$\Omega^\ast_{bc} \to \Omega_{bc}^{l}$ 
& 0.502
& 0.29 $\pm$ 0.03 \\
\hline 
$\Omega^\ast_{bc} \to \Omega_{bc}^{h}$ 
& 0.31 $\times$ 10$^{-2}$ 
& (0.01 $\pm$ 0.01) $\times$ 10$^{-4}$ \\
\hline 
\end{tabular}
\end{center}
\end{table}  

\end{document}